\documentclass[sigconf,nonacm,authorversion]{acmart}

\renewcommand\footnotetextcopyrightpermission[1]{} 

\usepackage{comment}
\usepackage[ruled,vlined,linesnumbered]{algorithm2e}
\usepackage{multicol}
\usepackage{url}

\usepackage[english]{babel}
\usepackage{blindtext}
\usepackage{ulem}
\usepackage{enumitem}
\usepackage{xcolor}
\usepackage{tikz}
\usepackage{caption}
\usepackage{subcaption}
\usepackage{array}
\usepackage{amsmath}
\usepackage{multirow}
\usepackage{float}
\usepackage{makecell}
\usepackage{balance}
\usepackage{enumitem}
\usepackage{titlesec}
\usetikzlibrary{fit,calc}
\setlist[itemize]{topsep=0pt, partopsep=0pt, parsep=0pt, itemsep=0pt}
\newcommand{\boldpara}[1]{\noindent \textbf{#1}}
\newcommand{\sys}{\emph{HiDe}\xspace}
\newcommand{\iperf}{\emph{iperf3}\xspace}

\newcommand{\longdistance}{\emph{long-distance}\xspace}

\newcommand{\remove}[1]{}

\renewcommand\footnotetextcopyrightpermission[1]{} 
\setcopyright{none}
\hypersetup{hidelinks}
\microtypecontext{spacing=nonfrench}
\acmDOI{}
\acmISBN{}
\acmConference[Submitted for review]{}
\acmYear{2025}
\copyrightyear{}
\acmPrice{}

\begin{document}

\date{}

\title{Data-Plane Telemetry to Mitigate Long-Distance BGP Hijacks}

\author{Satadal Sengupta}
\affiliation{Princeton University\country{}}
\email{satadals@princeton.edu}

\author{Hyojoon Kim}
\affiliation{University of Virginia\country{}}
\email{tcr5zr@virginia.edu}

\author{Daniel Jubas}
\affiliation{Five Rings LLC\country{}}
\authornote{Work done as a Master's student at Princeton University.}
\email{danieljubas@gmail.com}

\author{Maria Apostolaki}
\affiliation{Princeton University\country{}}
\email{apostolaki@princeton.edu}

\author{Jennifer Rexford}
\affiliation{Princeton University\country{}}
\email{jrex@princeton.edu}

\begin{abstract}
\noindent 
Poor security of Internet routing enables adversaries to divert user data through unintended infrastructures (hijack).
Of particular concern---and the focus of this paper---are cases where attackers reroute domestic traffic through foreign countries, exposing it to surveillance, bypassing legal privacy protections, and posing national security threats.
Efforts to detect and mitigate such attacks have focused primarily on the control plane while data-plane signals remain largely overlooked.
In particular, change in propagation delay caused by rerouting offers a promising signal: the change is unavoidable and the increased propagation delay is directly observable from the affected networks.
In this paper, we explore the practicality of using delay variations for hijack detection, addressing two key questions:
(1) What coverage can this provide, given its heavy dependence on the geolocations of the sender, receiver, and adversary?
and (2) Can an always-on latency-based detection system be deployed without disrupting normal network operations?
We observe that for 86\% of victim–attacker country pairs in the world, mid-attack delays exceed pre-attack delays by at least 25\% in real deployments, making delay-based hijack detection promising. To demonstrate practicality, we design \sys, which reliably detects delay surges from long-distance hijacks at line rate. We measure \sys{}’s accuracy and false-positive rate on real-world data and validate it with ethically conducted hijacks.
\end{abstract}

\maketitle

\pagestyle{plain}

\section{Introduction}
\label{sec:introduction}

\noindent
BGP hijacks are a well-known threat, where attackers exploit the poor security of BGP---the Internet’s default routing protocol---to redirect traffic through their own infrastructure. These attacks are dangerous, allowing attackers to eavesdrop and steal sensitive information from unsuspecting users \cite{MyEtherWallet,visahijack,KlaySwap,83,celer, italian_hijack,taiwan_hijack,youtube_hijack}.
Despite years of research aimed at addressing this issue, BGP hijacking continues to pose a significant threat~\cite{robachevsky2019routing}. 
Solutions like BGPsec~\cite{bgpsec} require ubiquitous adoption to be truly effective, which is a significant challenge given the decentralized nature of the Internet. On the other hand, mechanisms like RPKI~\cite{rpki}, while beneficial, are not bulletproof against all types of attacks.

Of particular concern, and the focus of this paper, are hijacks where domestic traffic is rerouted through a foreign location before reaching its intended destination, i.e., \emph{long-distance interception attacks}. Such attacks are especially troubling because they---unbeknownst to the user---expose the user's traffic to different jurisdictions and, consequently, to different privacy and surveillance laws. These rerouting incidents have significant implications~\cite{goldberg2014taking,goodin2018strange,york2014bgp}.
Yet, existing monitoring solutions~\cite{T=thousandeyes,
shi2012detecting,li2012buddyguard,sermpezis2018artemis} rely almost exclusively on control-plane signals. Concretely, they aim at finding anomalies in BGP route updates and are thus limited to what is visible from their vantage points or monitors~\cite{birgelee2019sico,10078883,birge2024global,york2014bgp,milolidakis2023effectiveness}.
Such an approach is also constrained by the slow convergence of BGP that can take minutes~\cite{holterbach2019blink}. 

In this paper, we investigate the usefulness of propagation delay, a data-plane signal, in detecting BGP hijacks and in particular \longdistance hijacks. Propagation delay is a promising yet underexplored signal for detecting such long-distance attacks, especially compared to well-studied control-plane signals. Unlike control-plane signals, propagation delay cannot be hidden from the victim, and in the case of \longdistance hijacks, the increase in delay is significant.
 For instance, for a UK-based source-destination pair, the traffic must cover an additional $15,025$ kilometers at least to travel via North Korea, causing a minimum additional round-trip time (RTT) of $75$ ms\footnote{This example assumes the speed of data transmission to be the speed of light in optical fiber, given by $c_f = 2c/3$ (approx.), where $c$ is the speed of light in vacuum~\cite{levin2015alibi}. Hereafter, in this paper, \emph{speed of light} refers to $c_f$, which is approx. $200$ km per millisecond (more precisely, $199.86$ km/ms).}---a latency the victim will directly experience. A change in propagation delay is also immediately observable, enabling a faster detection and mitigation opportunity compared to existing control-plane approaches.

However, building an effective delay-based BGP hijack detector is challenging. First, the expected increase in propagation delay---which is at the heart of a delay-based approach---is highly \textit{location-dependent}.
For example, a delay-based detector might successfully flag traffic between hosts in the U.S. being diverted through the UK but may fail to do the same if traffic is diverted through Canada. 
Additionally, delays can be caused by many other factors such as network congestion, host processing times, and noise in access networks. 
Second, even if we accurately detect long-distance BGP hijacks, doing so in a scalable, real-time manner remains challenging. Per-packet round-trip time calculation is expensive at line rate. Furthermore, maintaining state and monitoring every flow is intractable.
Thus, we pose the following research questions: \textit{(1) What fraction of possible BGP hijacks can a delay-based approach detect? (2) Can we design a practical, always-on monitoring system to detect and mitigate hijacks in a scalable manner without excessive cost?}

To these ends, we design \sys, a practical system for detecting hijacks using propagation delay, which relies on three key insights.
First, BGP hijacks occur at the IP-prefix level and affect all traffic routed to the targeted prefix, meaning that during a real hijack, no packet to the targeted prefix can have a delay less than the minimum required for the attacker’s route. By passively measuring the delays experienced by as many packets as possible and relying on the minimum per prefix, \sys can reliably and scalably detect spurious delay surges. 
Second, we observe that a BGP hijack induces a distinct pattern in the \emph{denoised} delay over time, clearly differentiating it from other events, such as congestion. Concretely, a hijack causes a \emph{sharp} surge with location-dependent but calculable \emph{minimum height}, which one can detect using a changepoint detection algorithm. Third, implementing \sys can be made practical by deploying it on high-speed programmable switches. This is possible, despite the rigid computation and memory constraints of such devices, thanks to our switch-native implementation of changepoint detection and scalable latency measurements.

Our comprehensive evaluation demonstrates that \sys is highly reliable (zero false negatives by design), minimally disruptive to real traffic, and implementable on commodity hardware. To assess \sys's effectiveness, we tested it against ethically-conducted real-world hijacks\footnote{Ethical issues are discussed in detail in our \emph{Ethics} section (Appendix~\ref{sec:ethics}).}
and found that it detects them within 0.5 second. Additionally, to evaluate its impact on regular operations, we run \sys on campus network traces (19 billion packets, 5.3~TB bytes). The results show that its combination of algorithms effectively minimizes false alarms (<$0.012\%$), even in the presence of highly noisy real-world RTT signals. Furthermore, \sys reduces the impact of such false alarms by identifying and correcting them within a median of 0.75 seconds without human intervention.
We implement \sys entirely on a programmable switch, showcasing its potential for seamless deployability on a network's border gateway with minimal hardware cost and no delay overhead to normal traffic\footnote{We will make our artifacts (analysis code, prototype, anonymized data, ethical hijack steps) publicly available upon acceptance (see Appendix~\ref{sec:appendix:open-science}).}.
\section{Background}

\noindent
In this section, we describe BGP hijacks, present our threat model, and explain the limitations of existing strategies.

\subsection{BGP-based attacks}
\noindent
BGP is the primary protocol that connects Autonomous Systems (ASes) by enabling them to exchange and forward route announcements for IP prefixes. Each AS advertises routes for the prefixes it owns, including an AS path indicating the sequence of ASes to traverse to reach it. Routers independently select the best route for each prefix based on attributes like path length and routing policies.

\boldpara{BGP hijacks.}
\noindent
A BGP hijack occurs when a malicious or compromised AS falsely advertises routes to IP prefixes it does not own or cannot reach, misleading other ASes into rerouting traffic through its infrastructure. Suppose $AS100$ legitimately owns the IP prefix $1.1.1.0/24$. A malicious AS, $AS200$, falsely announces ownership of $1.1.1.0/24$ to its BGP peers. These peers may accept the announcement as valid and propagate it to their own peers, spreading the false route across the network. As a result, traffic destined for $1.1.1.0/24$---for instance, originating from another prefix like $2.2.2.0/24$ owned by $AS300$---may get misrouted to $AS200$ instead of reaching $AS100$. This enables the attacker to eavesdrop on, fingerprint, manipulate, or drop this illegitimately-obtained traffic.
In some cases, the attacker may even serve malicious content by impersonating the legitimate destination (e.g., by rapidly acquiring a ``valid'' certificate first by exploiting the weakness in the certificate issuance verification process~\cite{birge2018bamboozling}).

\boldpara{BGP interception attack.}
\noindent
A BGP \textit{interception attack} (Figure~\ref{fig:interception-attack}) is a specific type of BGP hijack where the attacker intercepts traffic but forwards it to the original destination, enabling analysis or manipulation while remaining undetected by the end hosts. Of particular concern is a stealthy subset of these attacks where a sophisticated attacker employs techniques like \emph{AS-path poisoning} and the manipulation of \emph{BGP communities} to limit the propagation of malicious announcements in a bid to evade detection by BGP monitors near the victim. For instance, in the example above, $AS200$ could manipulate its announcements to propagate only to routers near $AS300$ while suppressing those to routers near $AS100$. This could cause traffic from $2.2.2.0/24$ to $1.1.1.0/24$ to get misrouted via $AS200$, while $AS100$ remains unaware as BGP monitors near it never observe the malicious route. 
Such an attack has been demonstrated by Birge-Lee et al.~\cite{birgelee2019sico}, which we ethically reproduce in \S\ref{subsec:setup-active}.

\subsection{Threat model}
\label{subsec:threat-model}

\begin{figure}[t]
\centering
\includegraphics[scale=0.55]{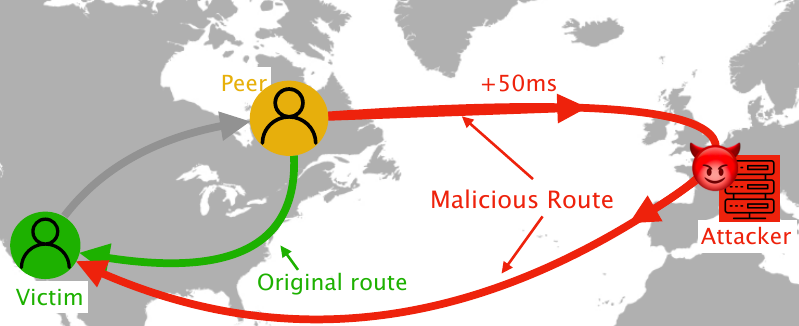}
\caption{An attacker in the UK exploits the lack of routing security to redirect traffic from a \emph{peer} host in the US---originally destined for a \emph{victim} host in the US---through the attacker’s own infrastructure. The \emph{mid-attack} path (in red) from the peer to the victim is longer than the original \emph{pre-attack} path (in green), adding an extra 50 ms of propagation delay.}
\label{fig:interception-attack}
\end{figure}

\noindent
We consider an adversary performing a \emph{stealthy} BGP interception attack---such as the one described above---to reroute traffic destined for a victim through distant infrastructure before forwarding it back to the victim.
This infrastructure is located in a faraway country, potentially under different privacy and security laws. The adversary is sophisticated, aware of detection systems, and employs evasion techniques to suppress forged advertisements~\cite{birgelee2019sico}. By rerouting traffic back to the victim, the attacker keeps connections alive, enabling traffic analysis while evading detection at the application layer. Such attacks can serve as tools in cyber warfare or surveillance.
Real-world examples of long-distance interceptions include (among numerous others) the rerouting of US-based traffic via the UK to enable surveillance (Figure~\ref{fig:interception-attack})~\cite{goldberg2014taking}, rerouting of US-based traffic managed by China Telecom via China undetected over $2.5$ years~\cite{goodin2018strange}, and rerouting of traffic between two hosts in Denver, USA via Iceland~\cite{york2014bgp}.

\subsection{Limitations of existing approaches}
\label{subsec:limitations-existing}

\boldpara{Protocol enhancement-based approaches are expensive or inadequately deployed.}
Multiple protocols have been proposed to secure routing on the Internet using cryptography to validate routing paths, including BGPSec~\cite{bgpsec}, RPKI~\cite{rpki}, and even future Internet architectures such as SCION~\cite{scion}. 
The key drawback of such solutions is that they require widespread deployment to be effective, which is hindered by the lack of incentives among independent parties. 
BGPSec requires online cryptographic signing and validation, which incurs high overhead, hindering adoption. RPKI~\cite{rpki} is less expensive and has seen more deployment, but only supports origin validation and not path validation. This makes RPKI unable to detect BGP interception attacks, which maintain the origin of the prefix (to allow traffic to reach its destination). Similarly, SCION requires worldwide deployment, extra infrastructure, and processing overhead on end-host software.

\boldpara{Control plane-based monitoring approaches are slow and susceptible to evasion.}
There is an abundance of research and industrial tools for detecting BGP hijacks~\cite{shi2012detecting,li2012buddyguard,sermpezis2018artemis, T=thousandeyes}. These approaches analyze the BGP route advertisements that are captured by monitors in different locations on the Internet to find anomalies. 
However, the detection may take too long. 
As BGP advertisements can take a long time to propagate to all locations, the detection speed depends on the location of the monitor and its distance from the victim. This can critically delay the reaction to a hijack, allowing the attacker enough time to achieve their goal.
Furthermore, while effective against hijacks caused by human error, such mitigation strategies can be evaded by sophisticated attackers. For example, previous works show that it is possible to avoid the monitors by carefully manipulating BGP communities to localize advertisements~\cite{birgelee2019sico, milolidakis2023effectiveness}.

\boldpara{Data-plane approaches lack practicality in real-world deployments.}
While some RTT-based detection methods exist, they fall short in production. 
Dart~\cite{sengupta2022continuous} introduces real-time RTT measurement in the data plane to address scalability challenges. While it employs min-filtering over windows and thresholds in an interception attack experiment, it cannot differentiate natural RTT variations (e.g., due to congestion), from malicious diversions---essential for effective hijack detection.
Oscilloscope provides a more advanced methodology for hijack detection~\cite{buhler2023oscilloscope}, but is based on idealized conditions, since it is evaluated only on emulated data. Even under these controlled scenarios, it suffers from high false-negative and false-positive rates.
Moreover, it does not operate directly in the data plane hardware, further limiting its applicability in production.
\section{Feasibility of delay-based detection}
\label{sec:feasibility}

\noindent
\sys{} relies primarily on propagation delay for real-time BGP hijack mitigation.
In this section, we examine the \textit{feasibility} of using propagation delay alone to defend against long‐distance interception attacks. In particular, we consider \emph{cross‐country} attacks---where both the victim and its peer reside in a \emph{victim country} $C_V$, while the attacker operates from a different \emph{threat country} $C_T$. Such attacks are common in nation‐state cyber warfare and surveillance.

\begin{figure*}[t]
     \centering
     \begin{subfigure}[b]{0.32\textwidth}
         \centering
         \includegraphics[width=0.95\textwidth]{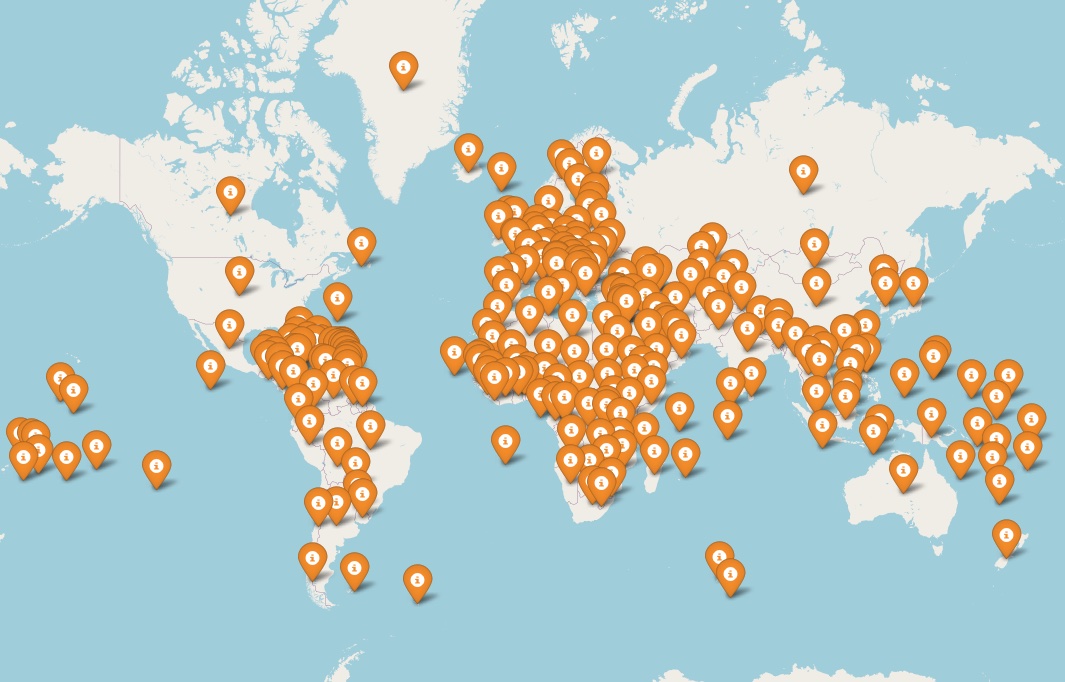}
         \caption{Centroids (orange icons) of the mainlands of 258 countries in the Natural Earth Admin-0 dataset~\cite{data2011natural}.}
         \label{subfig:ne-countries}
     \end{subfigure}
     \hfill
     \begin{subfigure}[b]{0.32\textwidth}
         \centering
         \includegraphics[width=0.95\textwidth]{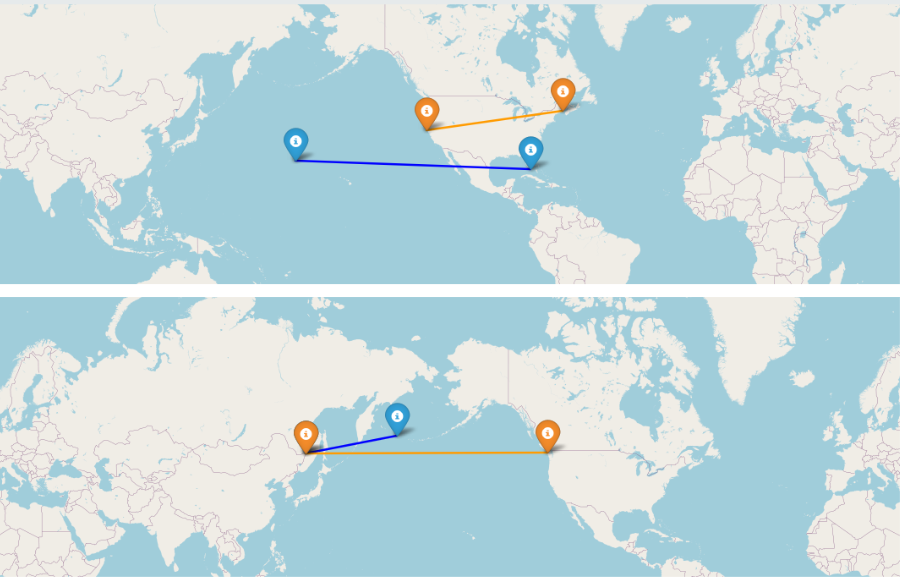}
         \caption{Top: Max. distance within country. Bottom: Min. distance between countries. (Blue: Entire country, orange: mainland).}
         \label{subfig:minmax-dist}
     \end{subfigure}
     \hfill
     \begin{subfigure}[b]{0.32\textwidth}
         \centering
         \includegraphics[scale=0.2]{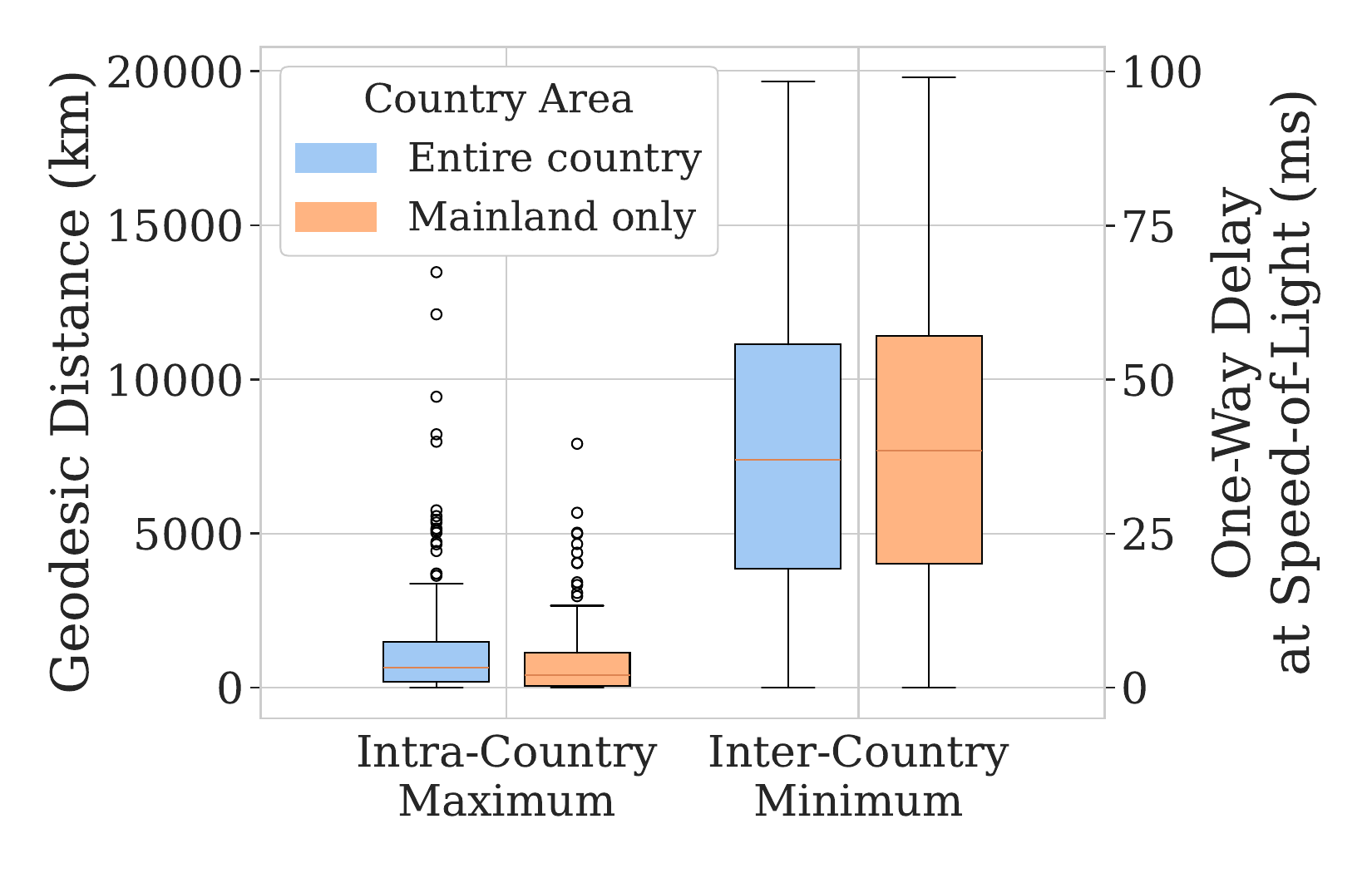}
         \caption{Distribution of max. within and min. between countries: distance (left y-axis) and one-way delay at $c_f$ (right y-axis).}
         \label{subfig:minmax-dist-owd}
     \end{subfigure}
        \caption{For all 258 countries (Figure a)---using both entire country areas and mainlands only (Figure b)---we compute each country’s maximum internal distance and its minimum distance to every other country, then plot both distributions (Figure c). Typically, a country’s foreign neighbors are more distant than its own farthest points.}
        \label{fig:country-distances}
\end{figure*}
\begin{figure*}[t]
     \centering
     \begin{subfigure}[b]{0.17\textwidth}
         \centering
         \includegraphics[height=0.8\textwidth]{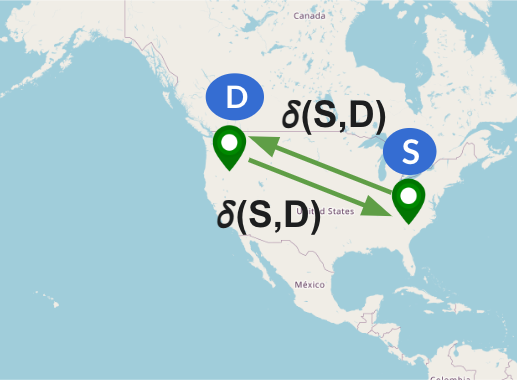}
         \caption{Distance before attack: $\delta_{pre} = 2\delta(S,D)$.}
         \label{subfig:pre-attack}
     \end{subfigure}
     \hfill
     \begin{subfigure}[b]{0.39\textwidth}
         \centering
         \includegraphics[height=0.35\textwidth]{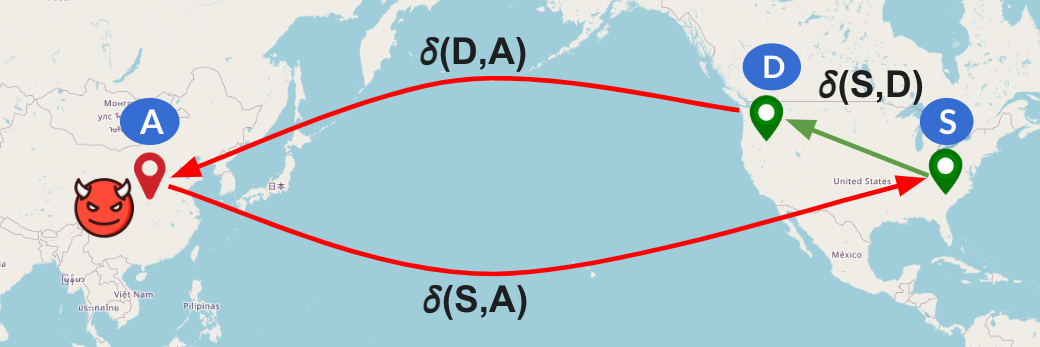}
         \caption{Distance in the middle of the stealthy interception attack:
         $\delta_{mid} = \delta(S,D)+\delta(S,A)+\delta(D,A)$.}
         \label{subfig:mid-attack}
     \end{subfigure}
     \hfill
     \begin{subfigure}[b]{0.39\textwidth}
         \centering
         \includegraphics[height=0.35\textwidth]{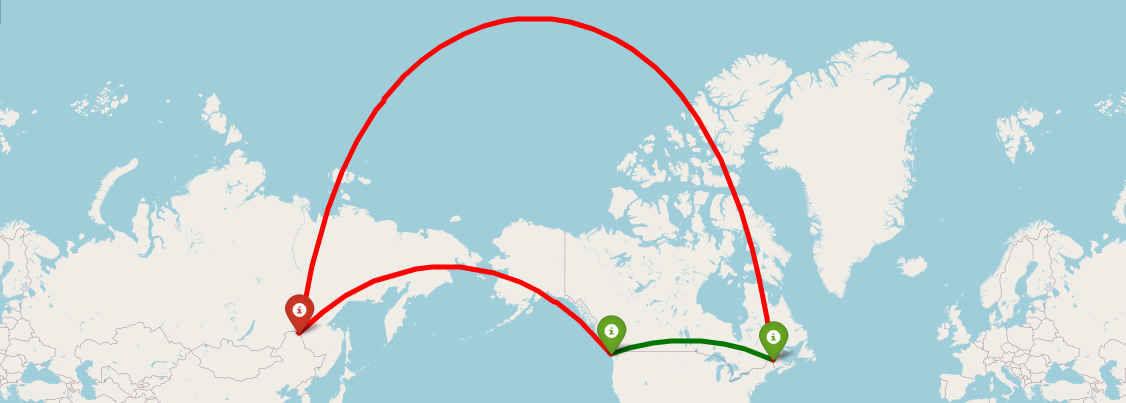}
         \caption{Locations of source, dest. in the US and attacker in China such that $\delta_{mid} - \delta_{pre}$ is minimized.}
         \label{subfig:optimal-attack}
     \end{subfigure}
        \caption{In this example, source S and destination D lie in mainland US and attacker A in mainland China. Figure (a) shows the \emph{pre-attack} round-trip distance $\delta_{pre}$ and (b) the \emph{mid-attack} round-trip distance $\delta_{mid}$, leading to the deviation $\delta_{deviation} = \delta_{mid} - \delta_{pre} = \delta(S,A)+\delta(D,A)-\delta(S,D)$. Figure (c) shows the \emph{most optimal attack} on the US from China, with curved lines indicating shortest great-circle paths. (Green: Original path, red: diversion due to attack.)}
        \label{fig:deviation}
\end{figure*}

\subsection{Key Questions and Observations}
\label{subsec:questions-observations}

\noindent
We ask the following questions about cross-country attacks:
\begin{enumerate}[leftmargin=*]
    \item Does traffic within a single country have significantly lower propagation delay than traffic across countries? If so, can we use this difference to detect cross‐country interception attacks?
    \item Are all countries equally \emph{defendable} using propagation delay-based detection? If some countries are more defendable, what geographic factors drive this difference?
    \item What fraction of cross‐country attacks can, in principle, be detected by comparing propagation delays---both under idealized (speed‐of‐light) assumptions and real-world measurements?
\end{enumerate}

\noindent
Our analysis in this section reveals the following:
\begin{enumerate}[leftmargin=*]
    \item Across 258 countries, the max. distance within a country (\emph{max. intra-country}) is typically far smaller than the min. distance from that country to any other country (\emph{min. inter-country}). The $25^{th}$/$50^{th}$/$75^{th}$ percentiles of max. intra‐country distances are 49 km, 413 km, and 1,129 km, respectively; the corresponding percentiles for min. inter‐country distances are 4,027 km, 7,689 km, and 11,420 km. Propagation delays follow a similar trend.
    \item Some countries are inherently more defendable using propagation delay-based detection than others. For example, Russia ranks among the least defendable, whereas New Zealand is one of the most defendable. More generally, larger countries with many nearby neighboring countries are less defendable, while smaller or more geographically isolated countries are more defendable.
    \item Considering the worst-case (least defendable) attack paths between every pair of countries, we find that 97\% cross-country attacks can be detected assuming speed-of-light RTTs, and 91\% and 86\% respectively, using real‐world measurements from two production datasets.
\end{enumerate}

\subsection{Intra- vs. Inter-Country Distances}
\label{subsec:country-distances}

\begin{figure*}[t]
     \centering
     \begin{subfigure}[b]{0.23\textwidth}
         \centering
         \includegraphics[scale=0.21]{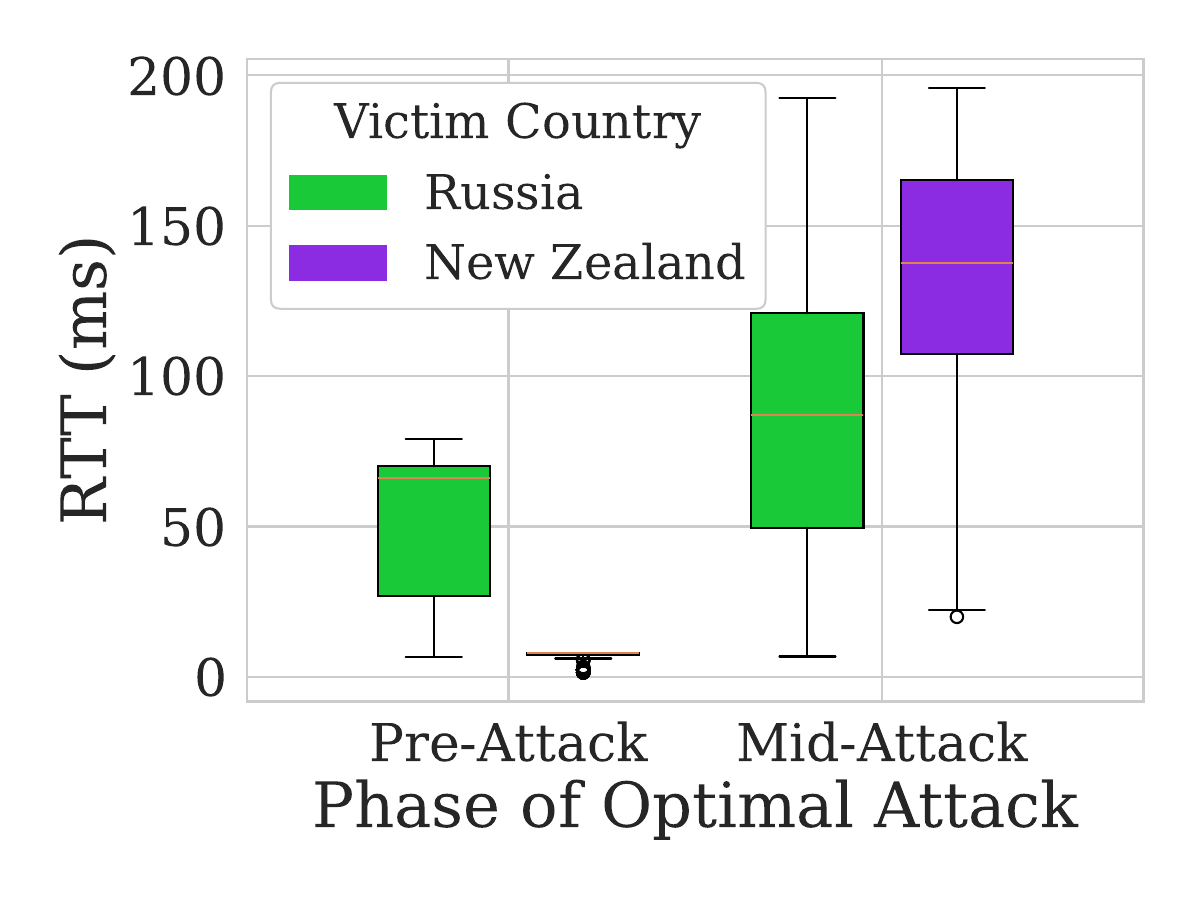}
         \caption{Pre- and mid-attack RTTs during all possible optimal attacks on 2 example countries.}
         \label{subfig:pre-mid-examples}
     \end{subfigure}
     \hfill
     \begin{subfigure}[b]{0.23\textwidth}
         \centering
         \includegraphics[scale=0.21]{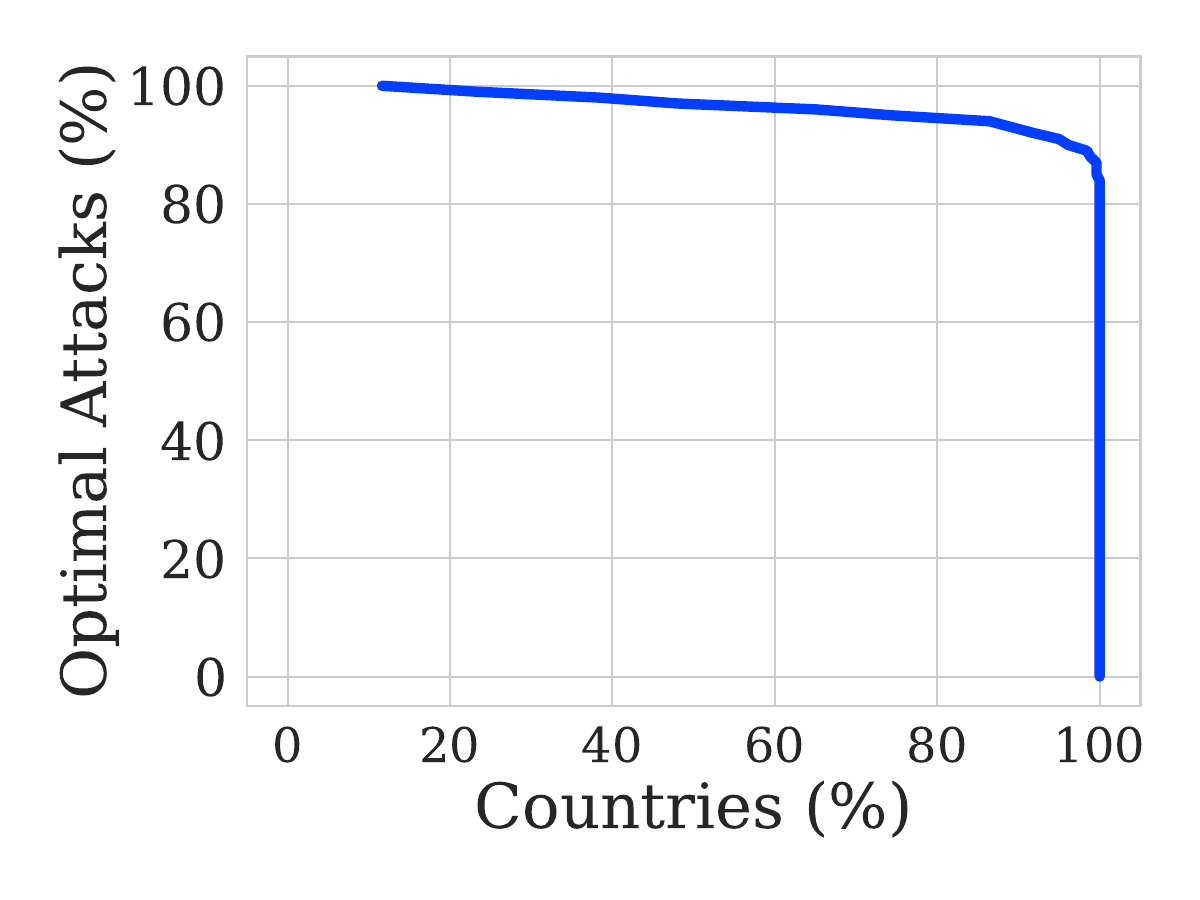}
         \caption{Percentage of countries vs. percentage of defendable optimal attacks.}
         \label{subfig:countries_attacks}
     \end{subfigure}
     \hfill
     \begin{subfigure}[b]{0.23\textwidth}
         \centering
         \includegraphics[scale=0.21]{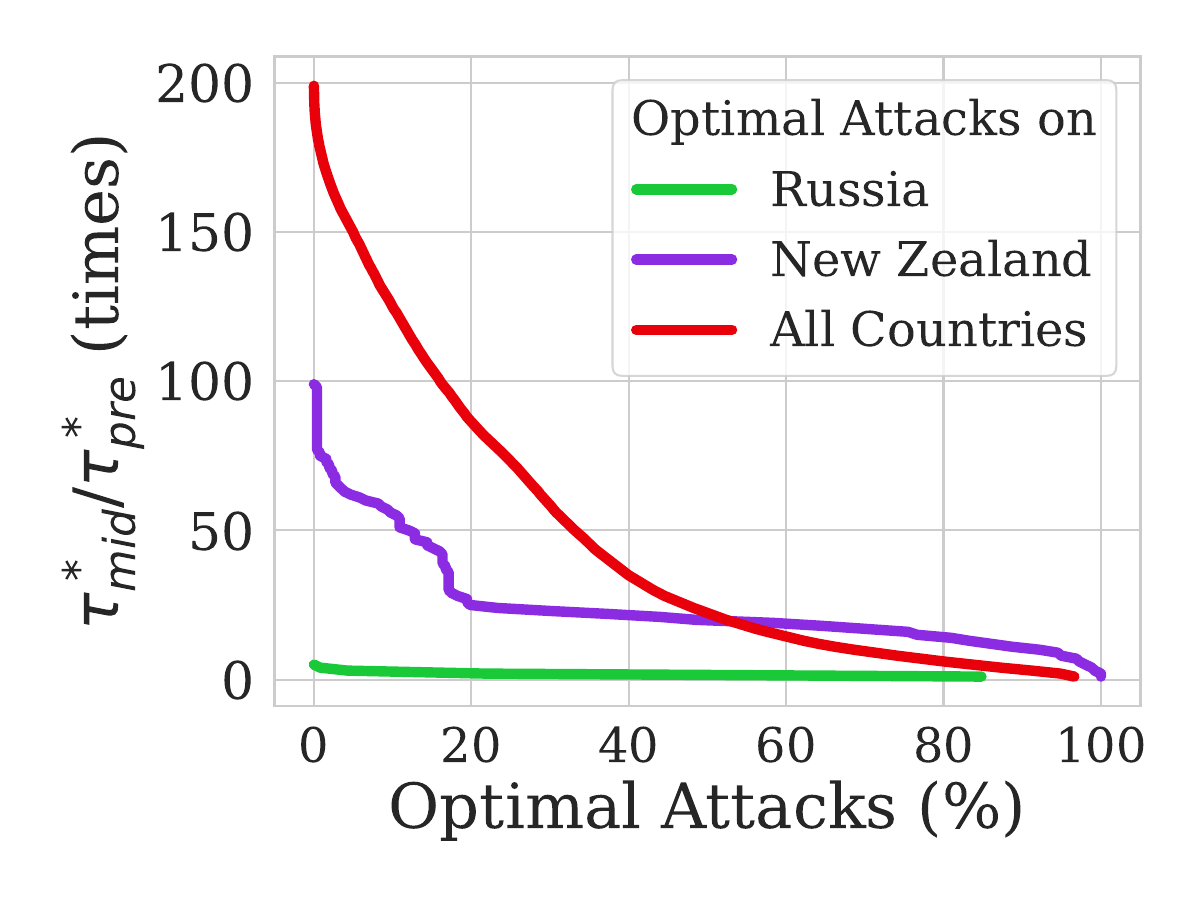}
         \caption{Attack coverage vs. mid-attack RTT (as a multiple of pre-attack RTT).}
         \label{subfig:mid_pre_times}
     \end{subfigure}
     \hfill
     \begin{subfigure}[b]{0.23\textwidth}
         \centering
         \includegraphics[scale=0.21]{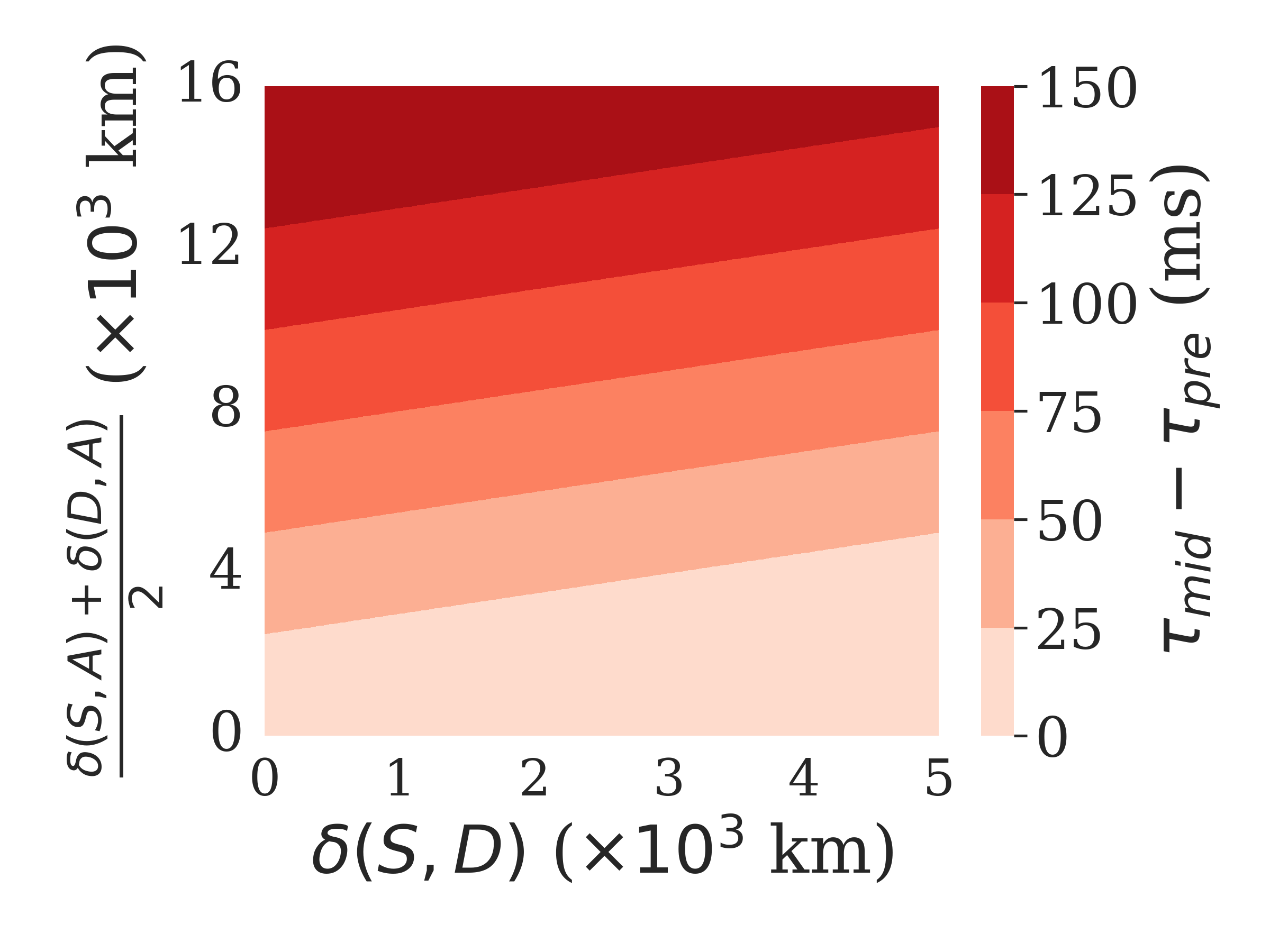}
         \caption{Deviation (z) as a function of dist. b/w hosts (x) and avg. dist. b/w hosts \& attacker (y).}
         \label{subfig:distance_deviation}
     \end{subfigure}
    \caption{Defendability against optimal attacks assuming speed-of-light RTT: With Russia and New Zealand (NZ) as example victim countries, (a) shows mid-attack RTT is typically much higher than pre-attack RTT; size and proximity of victim country to other countries determine the extent. Figure (b) shows that for 86\% countries can be defended against 94\% optimal attacks. Figure (c) shows Russia’s post-attack RTT peaks at 4x its pre-attack RTT (corresponding to 110 ms absolute difference), NZ at 100x (190 ms), and all countries combined at 198x (200 ms). Figure (d) shows that, when the victim and peer are co-located, the attacker must be 2,500 km away to induce a deviation of 25 ms; as the victim and peer separate, the attacker must be more distant to induce the same deviation.}
    \label{fig:optimal-attacks}
\end{figure*}

\boldpara{Goal.} Our first goal is to assess whether the great‐circle distance (shortest distance along Earth's curvature) between two hosts in the same country is significantly smaller than between hosts in different countries, so a cross‐country detour would induce a detectable increase in propagation delay.

\boldpara{Dataset.} We use the \emph{Natural Earth Admin-0 Countries} dataset---one of the most popular boundary datasets in the Geographic Information System community. It provides high‐resolution boundary coordinates (avg. $2.5$ km) for 258 countries (Figure~\ref{subfig:ne-countries}) \cite{data2011natural}.
Since some countries consist of multiple disconnected regions (e.g., contiguous US plus Alaska and islands), we distinguish each country’s \emph{mainland} (largest contiguous landmass) from \emph{entire country} (all regions combined).

\boldpara{Method.} For each country (entire country and mainland), we calculate: (1) Max. intra-country distance: the largest pairwise great-circle distance among all boundary points of the country, and (2) Min. inter-country distance: the smallest great-circle distance from any point in this country to any point in another. Figure~\ref{subfig:minmax-dist} illustrates these distances within the US (top) and between the US and China (bottom).

\boldpara{Observations.} Figure~\ref{subfig:minmax-dist-owd} shows: (1) min. inter-country distances far exceed max. intra-country distances, and (2) these distances for entire countries vs. mainlands are similar: we proceed with mainlands in the rest of this paper for better interpretability. At the speed of light, the $25^{th}$/$50^{th}$/$75^{th}$ percentile max. intra-country one-way delay (OWD) are 0.2 ms, 2.1 ms, and 5.6 ms, respectively; the corresponding values for min. inter‐country OWD are 20 ms, 38 ms, and 57 ms.

\subsection{Identifying least defendable attacks}
\label{subsec:optimal-attacks}

\noindent
In this subsection, we describe how we identify the least defendable attack given a $C_V$ and a $C_T$; later, we analyze defendability under ideal (\S\ref{subsec:defendability-ideal}) and realistic (\S\ref{subsec:defendability-measured}) conditions. Figures~\ref{subfig:pre-attack},~\ref{subfig:mid-attack} show the paths before and during the attack: with source S and destination D in $C_V$ and attacker A in $C_T$, the round-trip distance changes from $\delta_{pre}=2\delta(S,D)$ to $\delta_{mid}=\delta(S,D)+\delta(S,A)+\delta(D,A)$, resulting in a deviation of $\delta_{deviation}=\delta(S,A)+\delta(D,A)-\delta(S,D)$. In the worst-case (least defendable) attack scenario, $\delta_{deviation}$ is minimized by having S and D on $C_V$'s border and as far apart from each other as possible, and A on $C_T$'s border and as close to S and D as possible---we call this an \emph{optimal attack}. Under such an attack, we denote the pre-, mid-attack distances, and deviation as $\delta_{pre}^*$, $\delta_{mid}^*$, and $\delta_{deviation}^*$, respectively, and the corresponding RTTs as $\tau_{pre}^*$, $\tau_{mid}^*$, and $\tau_{deviation}^*$. Figure~\ref{subfig:optimal-attack} shows the optimal attack from China on the US, as an example. We compute the optimal attack for each ordered pair of victim and threat countries (258 x 257 attack scenarios).

\subsection{Defendability under ideal conditions}
\label{subsec:defendability-ideal}

\begin{figure*}[t]
     \centering
     \begin{subfigure}[b]{0.23\textwidth}
         \centering
         \includegraphics[scale=0.21]{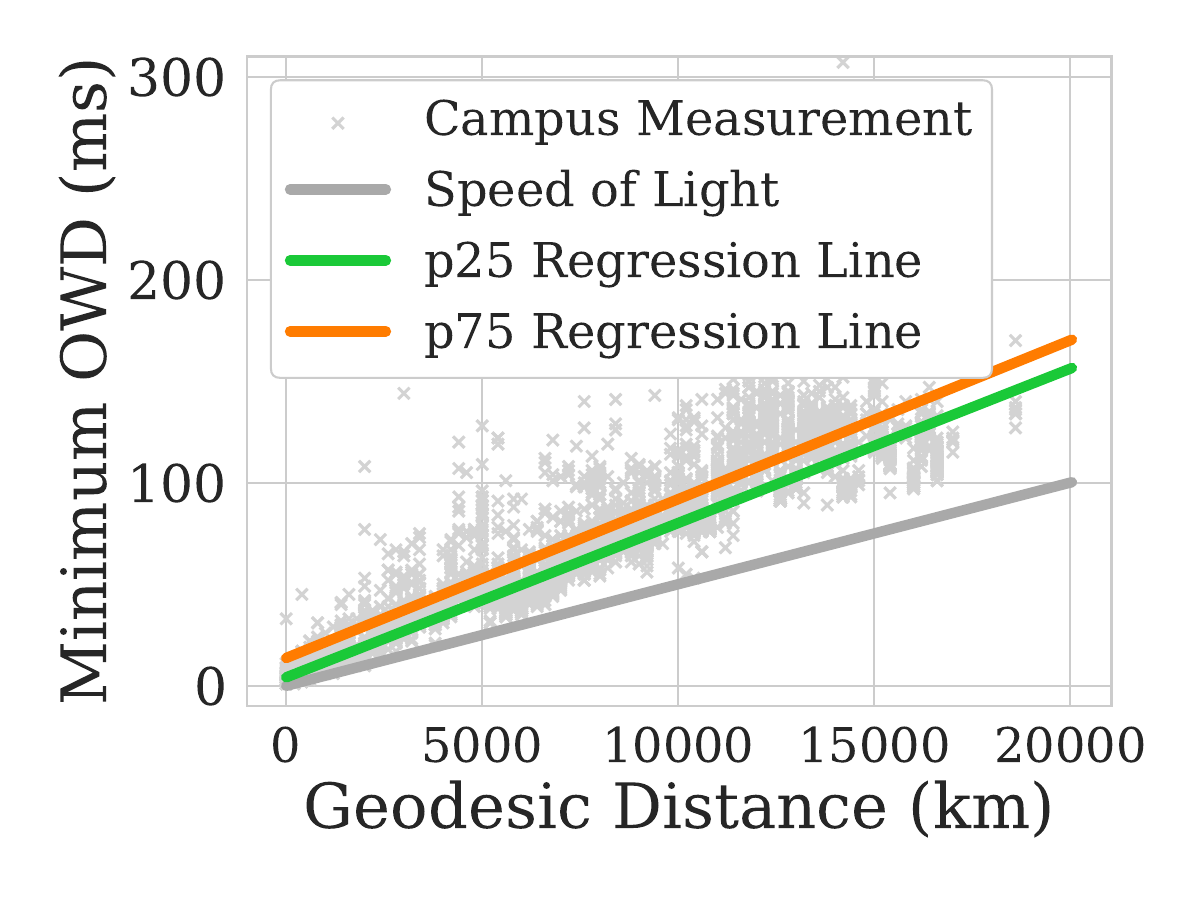}
         \caption{Campus measurements over 12 hours in May `22.}
         \label{subfig:campus-regression}
     \end{subfigure}
     \hfill
     \begin{subfigure}[b]{0.23\textwidth}
         \centering
         \includegraphics[scale=0.21]{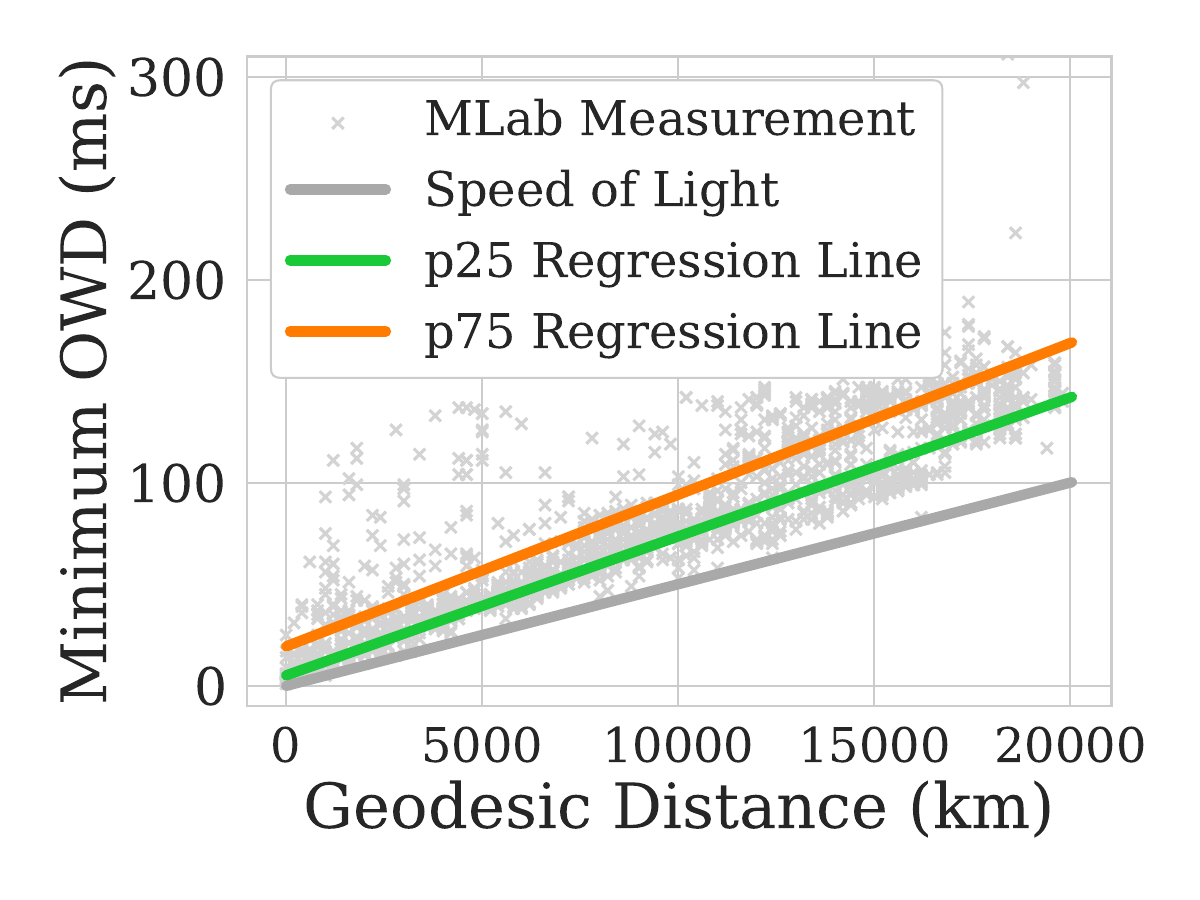}
         \caption{Google MLab measurements over 5 days in Dec. `24.}
         \label{subfig:mlab-regression}
     \end{subfigure}
     \hfill
     \begin{subfigure}[b]{0.23\textwidth}
         \centering
         \includegraphics[scale=0.21]{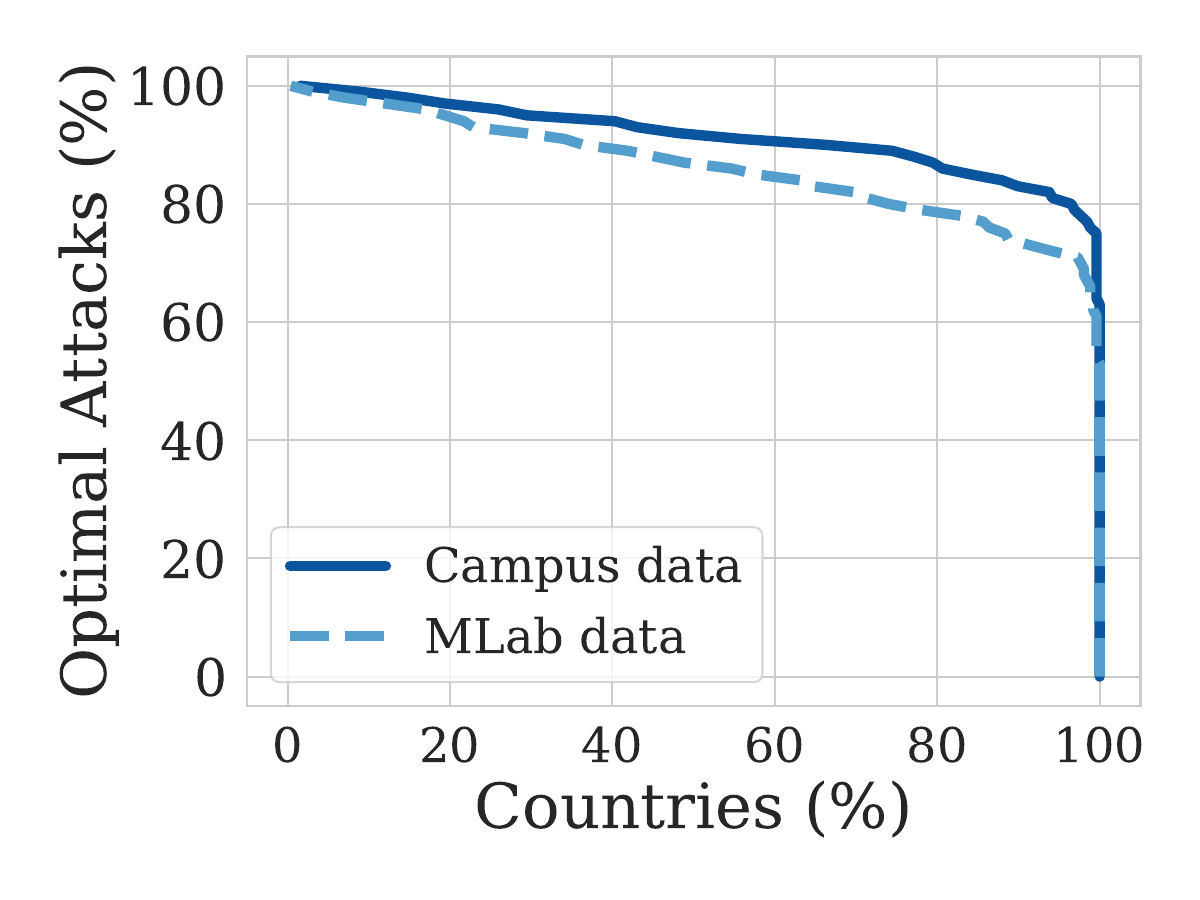}
         \caption{\% countries vs. \% defendable optimal attacks.}
         \label{subfig:countries-coverage-real}
     \end{subfigure}
     \hfill
     \begin{subfigure}[b]{0.23\textwidth}
         \centering
         \includegraphics[scale=0.21]{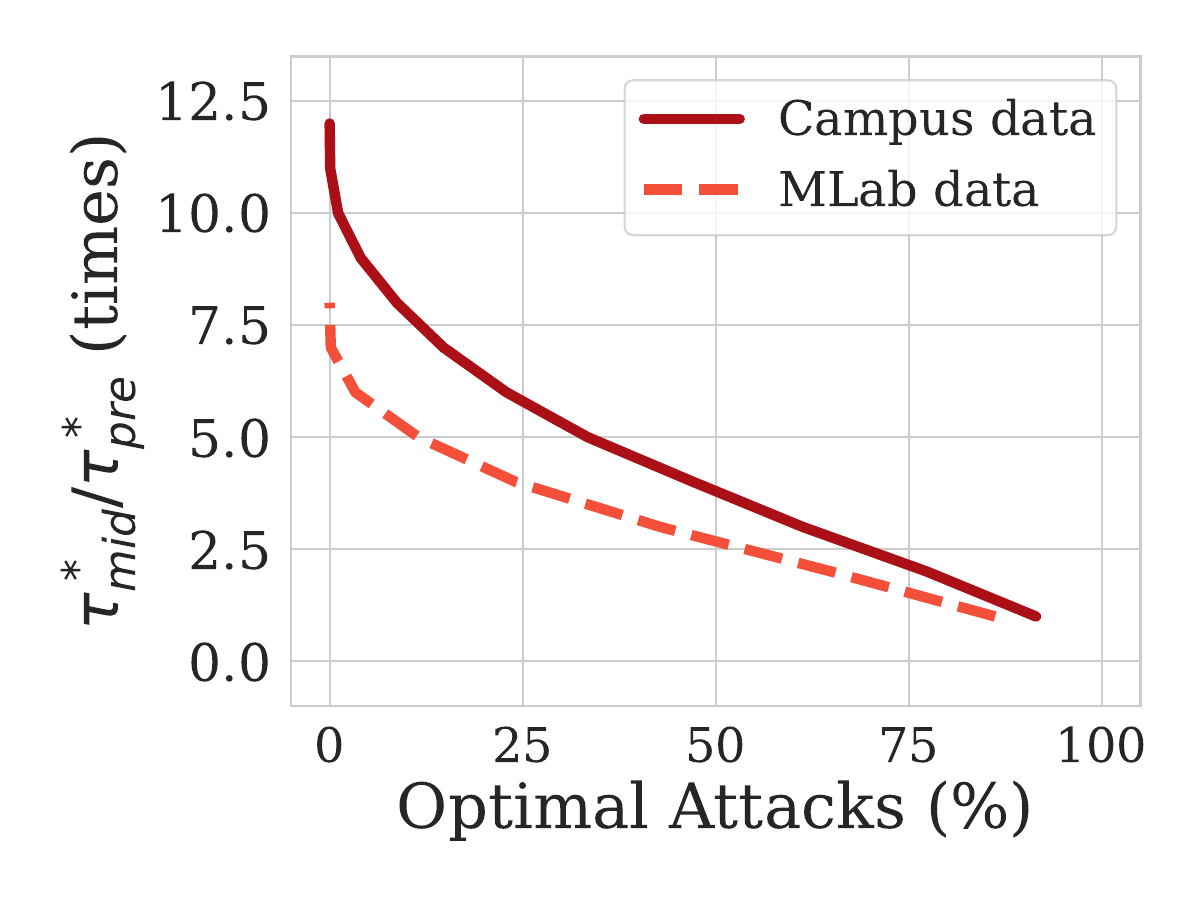}
         \caption{\% attacks vs. ratio of mid-attack to pre-attack RTT.}
         \label{subfig:attacks-times-real}
     \end{subfigure}
    \caption{Defendability against optimal attacks based on real measurements: We estimate (using linear regression) the $p25$ and $p75$ OWD for each 200 km distance bucket of our campus dataset and the Google MLab dataset, in (a) and (b) respectively. Using $p75$ OWD to estimate pre-attack and $p25$ to estimate mid-attack RTT, 85\% countries can be defended against 85\% attacks based on the campus dataset, and 85\% against 78\% based on MLab (Figure (c)). Figure (d) shows that the mid-attack RTT peaks at 12x pre-attack RTT in the campus dataset, and 7.5x in MLab.}
    \label{fig:measured-attacks}
\end{figure*}

\boldpara{Goal.}
In this subsection, we analyze the feasibility of detecting cross-country optimal attacks under \emph{ideal conditions}.

\boldpara{Assumptions.}
``Ideal conditions`` include: (1) Ideal network --- data travels at the speed of light, and (2) Ideal measurements --- our measurements capture the actual distance-based propagation delay. Under these assumptions, propagation delay = minRTT = RTT. Therefore, in this subsection, \emph{RTT} refers to propagation delay. We relax these assumptions in \S\ref{subsec:defendability-measured}.

\boldpara{Most and least defendable countries.}
Optimal attacks determine the lower bound of our defense capabilities. To assess defendability under optimal attacks, we compute propagation delay as the round-trip distance ($\delta_{pre}^*$ or $\delta_{mid}^*$) divided by speed of light. To illustrate the difference in defendability across countries, we select two examples: Russia, which has among the smallest median $\tau_{deviation}^*$, and New Zealand (NZ), which has one of the largest median $\tau_{deviation}^*$. Figure~\ref{subfig:pre-mid-examples} shows their pre-attack and mid-attack RTT distributions. NZ’s RTTs increase significantly mid-attack, making it more defendable; Russia’s RTT changes are smaller making it less defendable. In general, small countries with distant neighbors (low $\delta(S,D)$, high $\delta(S,A)+\delta(D,A)$) are the most defendable, while large countries with many close neighbors are the least defendable.

\boldpara{Attack coverage.}
To quantify defendability, we define \emph{attack coverage} as the percentage of optimal attacks that can be detected under some given condition. To compute overall coverage, we set the condition $\tau_{deviation}^*\ge5\,$ ms because: (1) 5 ms far exceeds typical noise in measurements (e.g., due to coarse-grained timestamps, rounding off errors, approximations in distance calculations, etc.), and (2) At speed of light, 5 ms corresponds to approx. 1,000 km of extra path length, so it captures meaningful geographic detours. The attack coverage for Russia (over 257 optimal attacks) is 85\% (the minimum for any country), while the same for NZ is 100\%. When expanded to all countries (i.e., 258 x 257 optimal attacks), the coverage is 96.6\%. Figure~\ref{subfig:countries_attacks} shows that 100\% (i.e., all) countries can be defended against 84\% attacks, 75\% against 95\%, 23\% against 99\%, and 11\% against 100\%. These results illustrate the promise and generality of propagation delay-based detection (in ideal conditions).

\boldpara{Attack coverage at given RTT deviations.}
To analyze the extent to which optimal attacks increase RTT in ideal conditions, we plot the attack coverage (x-axis) given the ratio between mid- and pre-attack RTT (Figure~\ref{subfig:mid_pre_times}) (absolute difference shown in appendix--Figure~\ref{fig:mid_pre_diff}). For NZ (purple), this ratio ranges from 1-100x (5-190 ms absolute difference), while for Russia (green), due to its large pre-attack RTTs, it is 1-4x (5-110 ms). For attacks on all countries (red), it is 1-198x (5-200 ms). The ratio is 2x (8 ms) for 95\% attack coverage on all countries, and 4x (23 ms) for 85\% coverage.

\boldpara{Deviation as a function of distances.}
The analysis of cross-country attacks does not inform us directly about the relationship between distance and RTT deviation. To bridge this gap, Figure~\ref{subfig:distance_deviation} shows RTT deviation (z-axis)---as a function of pre-attack distance ($\delta(S,D)$) (x-axis), and average distance between S-A and D-A (y-axis)---in a heatmap. The x- and y-axis are capped at the maximum intra- and minimum inter-country distances. The $85^{th}/95^{th}$ percentile pre-attack distances are 1,090 km and 1,872 km; to induce an RTT deviation of 25 ms, the attacker needs to be at an average distance of 3,045 km and 3,436 km, respectively, from S and D.

\subsection{Defendability \emph{in the wild}}
\label{subsec:defendability-measured}

\boldpara{Goal.}
While our observations under ideal conditions are promising, defendability may differ in the real world because: (1) the actual propagation delay may be larger than the speed-of-light RTT due to longer physical paths, and (2) RTT measurements may not capture the actual propagation delay due to congestion or poor channel conditions (in wireless networks). In this subsection, we evaluate defendability in realistic conditions using two production datasets.

\boldpara{Datasets.} Our datasets are outlined below:
\begin{itemize}[leftmargin=*]
    \item \textbf{Campus dataset:} We collect traffic from 7.5 M TCP flows on our campus over 12h on a weekday in May `22, and compute RTTs by matching data packets with ACKs~\cite{chen2020measuring}.
    \item \textbf{MLab dataset:} We collect minRTTs of 4.3 M TCP flows from NDT7-based measurements over 5 days in Dec. `24~\cite{mlab_minrtt}.
\end{itemize}
For each flow in each dataset, we collect geolocations of the source and destination.
We discard flows whose minRTT indicates shorter distances than those permitted by their reported geolocations, which is physically impossible. 
Finally, we group remaining measurements by source-destination /24 prefixes, because (1) a /24 prefix is the smallest unit on which a BGP hijack can be launched, and (2) aggregating by prefix improves the chance of measuring true minRTT (\S\ref{subsec:rtts-noise}).

\begin{figure*}
\centering
\begin{minipage}[t]{0.32\textwidth}
    \centering
    \includegraphics[width=.99\linewidth]{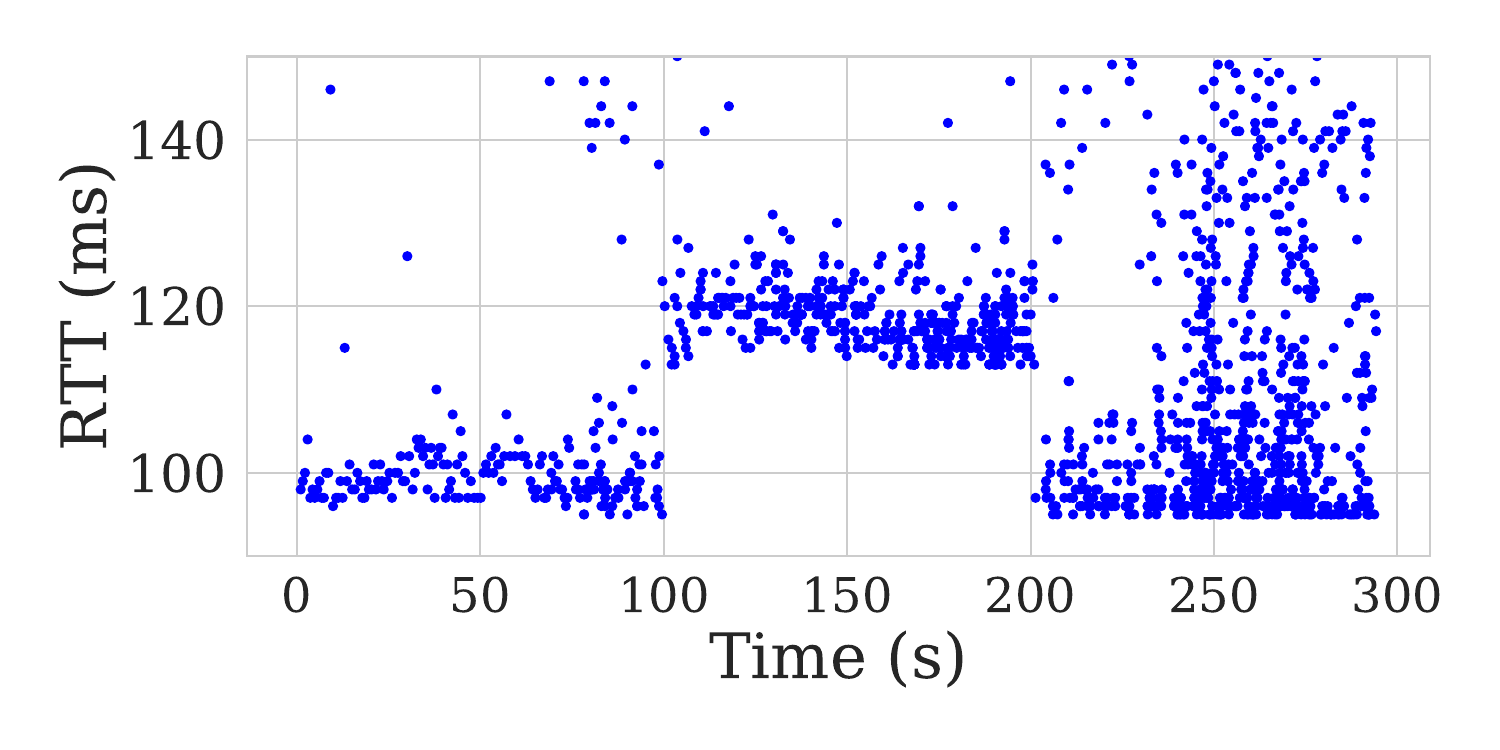}
    \captionof{figure}{\emph{Abrupt} and \emph{significant} rise and fall in RTT due to interception attack launched (ethically) at 100$^{th}$ second and withdrawn at 200$^{th}$.}
    \label{fig:bitcoin-attack-surge}
\end{minipage}
\hfill
\begin{minipage}[t]{0.32\textwidth}
    \centering
    \includegraphics[width=.99\linewidth]{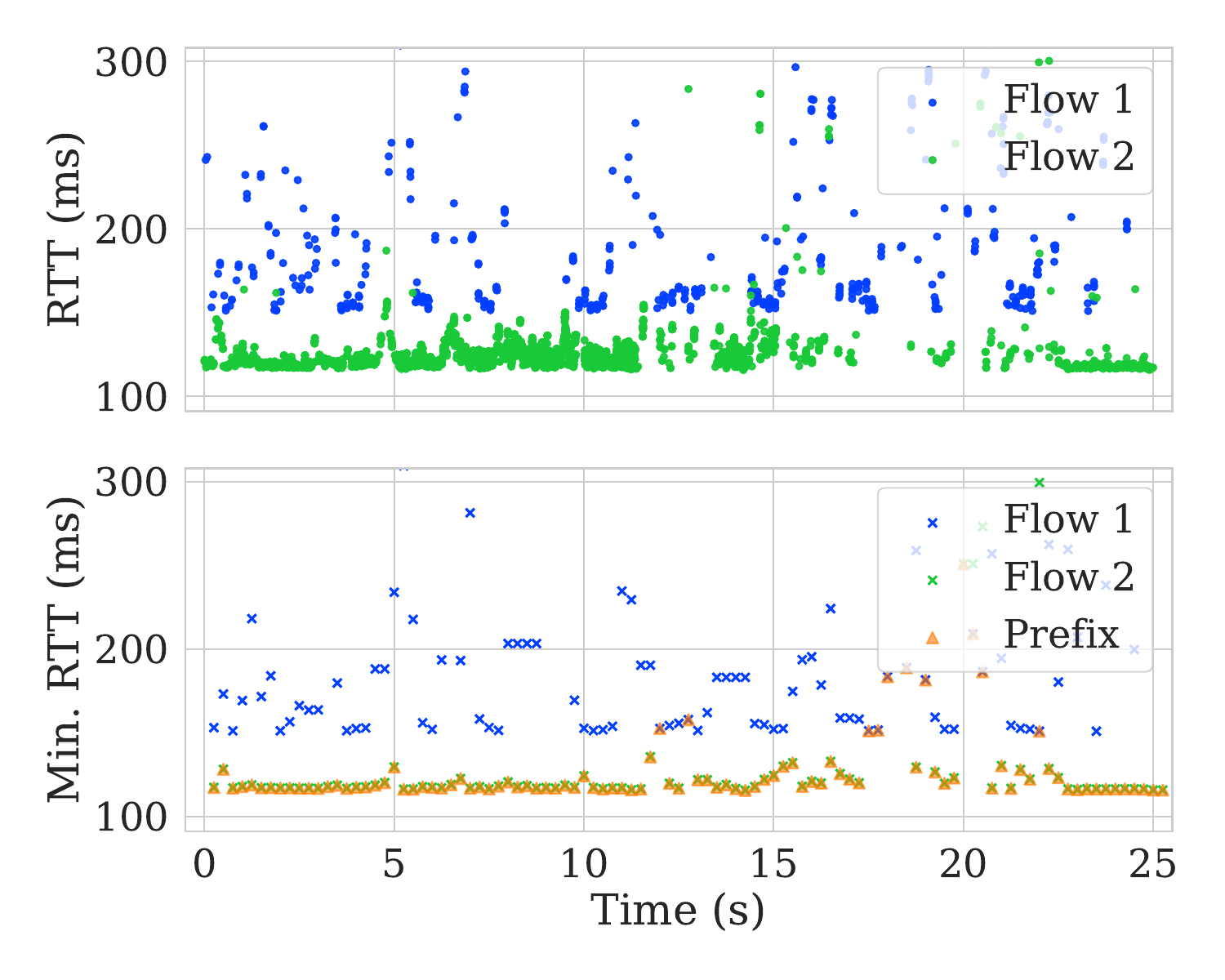}
    \captionof{figure}{Top: Flow 1 (blue) with noisy RTTs and flow 2 (green) with stable RTTs. Bottom: Aggregating by prefix stabilizes minRTTs (orange).}
    \label{fig:minrtt-prefix}
\end{minipage}
\hfill
\begin{minipage}[t]{0.32\textwidth}
    \centering
    \includegraphics[width=.99\linewidth]{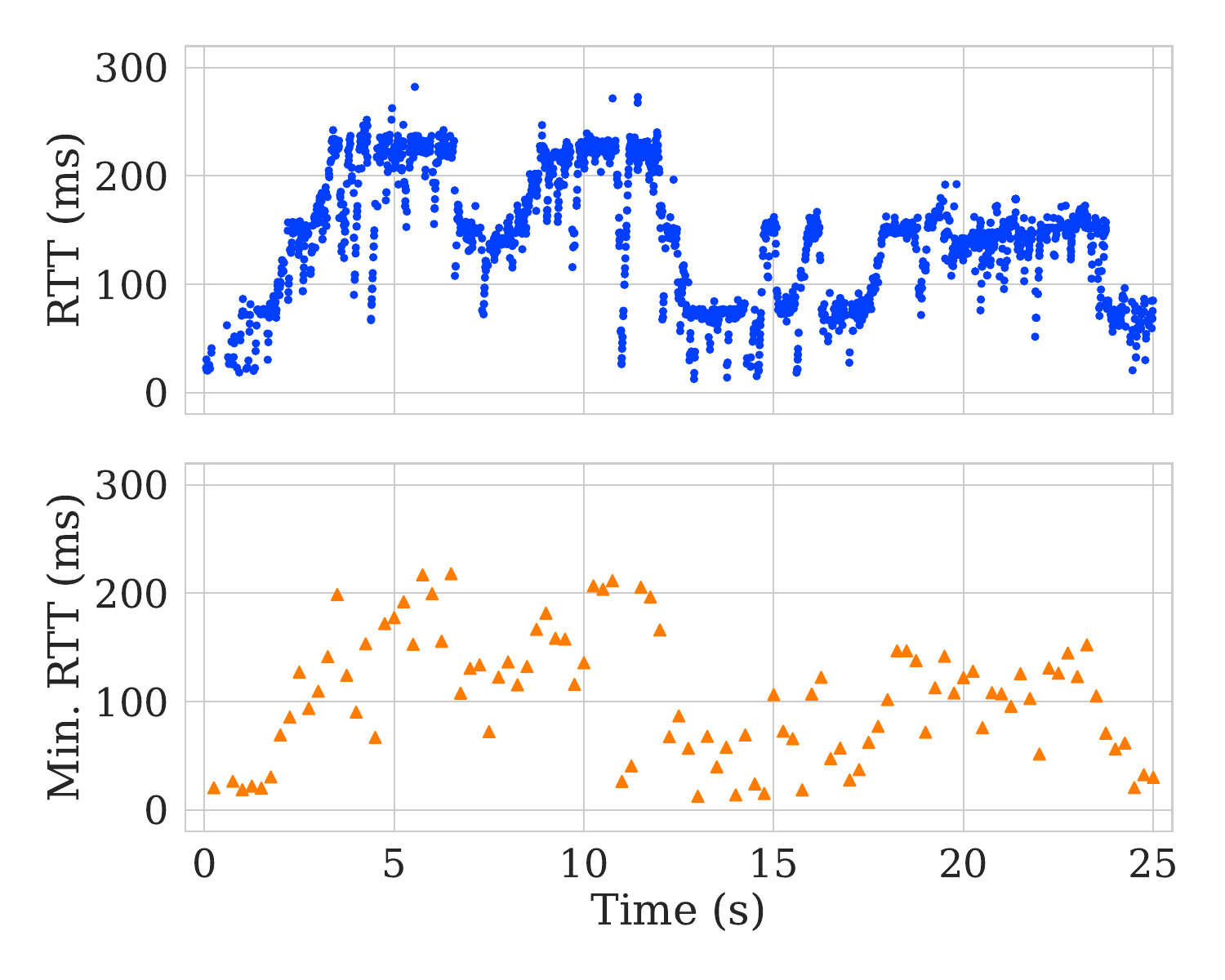}
    \captionof{figure}{Prefix with noisy RTTs (top) that produce noisy minRTTs (bottom) despite windowing. Such prefixes are less defendable.}
    \label{fig:rtt-noise}
\end{minipage}
\end{figure*}

\boldpara{Estimating propagation delay from distance.}
To quantify defendability in realistic settings, we estimate real‐world propagation delay between any two hosts from their great‐circle distance $d$. This real-world propagation delay may vary across host pairs separated by the same distance $d$ depending on: (1) Geolocation: Some regions in the world have denser network connectivity than others, (2) Routing policies: Some providers make shorter paths available than others, (3) Stable queues: Some paths experience consistent queuing delay due to deep buffers, etc. Furthermore, even if the true propagation delay is same, we may measure different minRTTs due to transient congestion. It is infeasible to collect reliable minRTTs from all possible attack locations in all 258 countries. Instead, we apply the following method to both our campus dataset (215 countries) and Google MLab (234 countries) to capture variability:
\begin{enumerate}[leftmargin=*]
    \item \textbf{Bin distances:} Divide all distances up to 20,075 km (Earth’s diameter) into 200 km bins ($\approx$1 ms at $c_f$).
    \item \textbf{Assign prefixes to bins:} For each source–destination prefix pair, compute its great‐circle distance, assign it to the appropriate bin, and record its minOWD (minRTT/2).
    \item \textbf{Percentile computation:} Within each bin, compute the $p$th percentile of these minOWDs for $p=1,\dots,100$.
    \item \textbf{Regression fitting:} Fit one linear regression per percentile across all bins (e.g., a $p$=1 line through every bin’s $1$‐percentile).
\end{enumerate}
Figures~\ref{subfig:campus-regression} and~\ref{subfig:mlab-regression} plot the $p$=25 and $p$=75 regression lines for the campus and MLab datasets, respectively. The campus data shows a narrower inter-quartile range---likely because the location of one end is fixed (on campus) and the entire variability is due to the remote host---whereas for MLab, the servers and clients are in different locations. With these delay estimates, we proceed to evaluate defendability against optimal attacks under realistic conditions.

\boldpara{Estimating pre- and mid-attack minRTTs.}
The variability of minOWDs for the same distance poses a challenge: if our detection is unlucky, it could measure a higher percentile minRTT before the attack and a lower percentile during the attack, causing the deviation in minRTT to be much lower than speed-of-light deviation. To model such a scenario, we use the upper quartile ($75^{th}$ percentile) minOWDs to estimate pre-attack minRTTs, and the lower quartile ($25^{th}$ percentile) minOWDs to estimate mid-attack minRTTs. Then, we evaluate defendability using the same metrics as before.

\boldpara{Observations.}
Using the condition $\tau_{deviation}^*\ge5\,$ms, the overall attack coverage is 91\% on the campus dataset and 86\% on MLab. Figure~\ref{subfig:countries-coverage-real} shows that in the campus data, 100\% countries can be defended against 63\% attacks, 75\% against 89\%, and 2\% against 100\%. In MLab, 100\% countries can be defended against 53\% attacks, 75\% against 80\%, and <1\% against 100\%. Real-world defendability is therefore less than in the ideal case---especially in MLab, where minRTT variability is higher. Figure~\ref{subfig:attacks-times-real} plots coverage versus the mid-/pre-attack minRTT ratio (absolute diff. in appendix--Figure~\ref{fig:cvg_diff_real}). The ratio ranges from 1–12x (5–290 ms) for campus; 1–7.5x (5–250 ms) for MLab. Appendix-figures~\ref{fig:campus_heatmap},~\ref{fig:mlab_heatmap} present corresponding distance-deviation heatmaps, which follow similar trends.

\subsection{Takeaways}
\label{subsec:defendability-takeaways}

\noindent
Our analysis shows that propagation‐delay measurements offer a highly effective signal to defend against cross‐country interception, primarily because diverted paths almost always incur substantially greater delays than pre-attack paths. 
Although real‐world variability degrades detection coverage compared to ideal conditions, our focus on worst‐case (optimal) attacks means these results are conservative---actual detection performance will often exceed our current estimates.
We believe our findings are sufficiently strong to justify designing an interception-detection based solely on propagation delay.
At the same time, a range of factors influences how accurately any given attack can be detected: the victim country’s size and distance from potential adversaries; the exact geolocations and separation of victim and peer hosts; the true lengths of the pre-attack and mid-attack network paths; transient or persistent congestion along those paths; and the precision of our measurement and aggregation techniques.
With these insights in mind, in the next section, we present \sys{}---a scalable, always-on, data-plane system for detecting and mitigating interception attacks.
\section{\sys{}: Overview}
\label{sec:overview}

\noindent
\sys is a BGP-interception mitigation system that runs entirely on a programmable switch and uses real-time minRTT measurements for detection and mitigation. In this section, we present the key insights that drive \sys.

\boldpara{Converting noisy RTT into a reliable detection signal.}
During a hijack, all traffic to a victim prefix must traverse the longer path via the attacker, so no RTT sample can be shorter than the minimum propagation delay via the attacker. \sys exploits this by passively collecting RTT samples for every TCP data‐ACK pair at the network border (thereby avoiding noise from the internal network), aggregating samples per prefix, and tracking the minRTT per time window.
By monitoring these minRTTs, \sys detects hijacks as sudden, sustained spikes in delay. For example (Figure~\ref{fig:bitcoin-attack-surge}), hijacking Bitcoin traffic ethically from a Stockholm client via Amsterdam causes the minRTT to jump by about 20 ms at attack start and to fall back when the hijack ends. A changepoint detection algorithm can reliably identify such shifts.

\boldpara{Prioritize guaranteed protection over broad coverage.}  
We posit that operators prefer a system that reliably defends a well‐defined subset of prefixes rather than a best‐effort approach that ``covers'' everything but floods them with false alarms. Section~\ref{subsec:profiling-prefixes} shows how \sys restricts its scope to prefixes it can protect with high confidence. When false positives do occur, \sys continues  measuring RTT and automatically rolls back its mitigation if the spike proves transient.

\boldpara{Optimize for commodity hardware.}  
\sys stores only per-prefix state---the running minRTT and sample count per time window---instead of expensive per-flow or per-packet state. It employs a lightweight, two-window, threshold-based changepoint detector that is hardware-amenable. On Tofino2, \sys uses native primitives (\emph{mirror} and \emph{packetgen}) to generate occasional packet replicas for RTT measurement, false-positive correction, and (optinal) user alerting---while forwarding all other traffic at line rate with zero additional latency.
\section{\sys: Methodology}
\label{sec:methodology}

\subsection{Compute location-based lower bound}
\label{subsec:physics-based}

\boldpara{Translating user input into geographic locations.}
The user provides \sys with the IP prefix of the home network and the threat regions they want to protect their data from, based on policy decisions or anticipation of threats. The threat regions are either names of countries or enclosed polygons of geographic coordinates. \sys also obtains from its data plane the destination prefixes observed by it. The user can, \emph{optionally}, set threat regions \emph{per destination prefix}. Eventually, the control plane converts all the information into triplets of geolocation information: \{source\_coordinates, destination\_coordinates, threat\_coordinates\_list\} using public geolocation services (\emph{IPinfo}, \emph{MaxMind}) and public geographic datasets (\emph{Natural Earth Admin-0})~\cite{ipinfo, maxmind, data2011natural}.

\boldpara{Computing lower bound of mid-attack RTT.}  
For each location triplet, we first identify the \emph{optimal attack}, i.e., the attacker's location in the threat region that minimizes mid-attack round-trip distance. Note that this distance can be much higher than in the optimal attacks in \S\ref{sec:feasibility} because there, source and destination were always on the victim country's border whereas here, they are almost always inland. Next, we compute the minimum possible mid-attack RTT for this optimal attack ($\tau_{mid}^*$), based on the speed of light. We designate this \emph{lower bound} RTT as the  \emph{absolute threshold} of our changepoint detector: \sys flags an attack whenever the observed minRTT reaches $\tau_{mid}^*$, guaranteeing \emph{zero false negatives} (see \S\ref{subsec:fp-impact} for false positives). While we could choose a less conservative bound---e.g., the $25^{th}$-percentile estimated latency in 200 km buckets (\S\ref{subsec:defendability-measured})---we opt for the most conservative threshold to \emph{guarantee} protection, at the expense of coverage, as is our design goal (\S\ref{sec:overview}).

\subsection{Reduce noise in the RTT signal}
\label{subsec:rtts-noise}

\boldpara{Aggregating by prefix to reduce impact of noisy flows.}
BGP attacks target prefixes, with a $/24$ prefix being the smallest possible target. All flows to an attacked prefix experience the same underlying change in propagation delay, but noisy RTTs in individual flows can obscure this change. We aggregate RTT samples by prefix before computing the minimum RTT per window (discussed next), as at least one flow per window is likely to produce a sample representative of the true propagation delay. Figure~\ref{fig:minrtt-prefix} demonstrates this for a US-based destination prefix with one noisy and one stable flow. Also, prefix-level aggregation reduces switch memory requirements from per-flow to per-prefix, which is significant.

\boldpara{Windowing to discard short-term fluctuations.}
We divide streams of per-prefix RTT samples into non-overlapping time windows of a fixed size (i.e., \emph{tumbling} windows) and compute the minRTT in each window. This helps filter out short-term spikes in RTT due to benign confounding factors like queuing delay from short-lived congestion, end-host processing delays, and TCP oddities like \emph{delayed ACKs}~\cite{sengupta2022continuous}. We select a sub-second (e.g., 0.25- or 0.5-second) time window---while minRTT can be measured more reliable by with longer time windows, it would delay mitigation allowing an attacker more time to complete their attack.
Finally, tracking minRTTs in \emph{non-overlapping} tumbling windows requires only per-flow state, as opposed to \emph{overlapping} sliding windows, making it more suitable for a switch implementation.

\subsection{Cover vulnerable \& defendable prefixes}
\label{subsec:profiling-prefixes}

\boldpara{Identifying vulnerable prefixes.}  
BGP interception attacks primarily target prefixes that host sensitive services---government sites, banking portals, cryptocurrency nodes, and the like---because such websites handle sensitive data from users around the world. The attacker places itself between the user and the server---intercepting ``valuable'' data. \sys prioritizes these vulnerable server prefixes for protection. By default, it excludes prefixes used exclusively by WiFi or cellular access networks---since they rarely host critical services---unless the operator explicitly includes them.

\boldpara{Identifying defendable prefixes.}
Some destination prefixes have RTTs that are \emph{consistently} noisy or high even under benign conditions, making it nearly impossible to detect interception attacks on them from certain threat regions without excessive false positives. For example, Figure~\ref{fig:rtt-noise} shows a prefix where the minRTT often exceeds 100 ms, making it impractical to defend against a threat region that causes a small deviation in comparison. Further analysis of such prefixes reveals that often, they tend to be associated with client-side access networks, such as cellular or WiFi, which \sys does not defend by default anyway.
Concretely, during a \emph{profiling} phase independent of the detection phase, we monitor the \emph{max. of min. RTTs} in tumbling time windows for each destination prefix. Later, we defend a prefix against a threat region only if
$\tau_{mid}^* - max(RTT_{min}) > \lambda$,
where $\lambda$ is called the \emph{surge threshold}. $\lambda$ defines the min. increase in $RTT_{min}$ required between two consecutive windows to flag an attack, and can be set to a constant (e.g., 10 ms) or a fraction of $max(RTT_{min})$ (e.g., 10\%). A lower $\lambda$ provides broader coverage but increases susceptibility to false positives, and vice-versa. Users can adjust $\lambda$ based on their desired trade-offs. We evaluate the false positive rate for different values of $\lambda$ in \S\ref{sec:evaluation}.

\subsection{Switch-amenable changepoint detection}
\label{subsec:changepoint-detection}

\noindent
We implement changepoint detection directly in switch hardware by combining the techniques described so far in an approach called the \emph{two-window algorithm}. This involves tracking the per-prefix min. RTT ($RTT_{min}^i$) for each tumbling window $i$. Once the $i^{th}$ window completes ($i > 0$), we compare $RTT_{min}^{i-1}$ and $RTT_{min}^{i}$ and mark the prefix as \emph{attacked} if both the following \emph{surge} conditions are met:\\
(1) $RTT_{min}^{i-1} < \tau_{mid}^*$ and $RTT_{min}^{i} > \tau_{mid}^*$: The minRTT crosses the absolute threshold between two consecutive windows.\\
(2) $RTT_{min}^{i} - RTT_{min}^{i-1} > \lambda$: The minimum RTT surges by at least the surge threshold in consecutive windows.

\boldpara{Valid windows.} Only time windows that contain at least 5 samples are considered valid and used for detection---windows with fewer samples do not necessarily benefit from min-filtering and could lead to false positives.

\subsection{Minimize impact of false positives}
\label{subsec:fp-impact}

\noindent
Despite reducing false positives, our detection algorithm is not foolproof and may occasionally generate them. To minimize their impact and to eliminate the need for human intervention, \sys employs an automatic \emph{false positive correction} mechanism. When an attack is detected, \sys blocks the affected prefix and simultaneously initiates active probing by sending ICMP echo packets to the most recently active IP address in the prefix at each time window. It monitors the corresponding RTT, and if the RTT falls below $\tau_{mid}^*$, the prefix is unblocked, and detection resumes, minimizing disruption to regular operations. To limit probe traffic, \sys reduces the probe rate to one per minute after five minutes of attempts. These parameters are user-adjustable for flexibility.
\section{\sys: System}
\label{subsec:system-description}

\begin{figure*}[t]
\centering
\includegraphics[width=.72\linewidth]{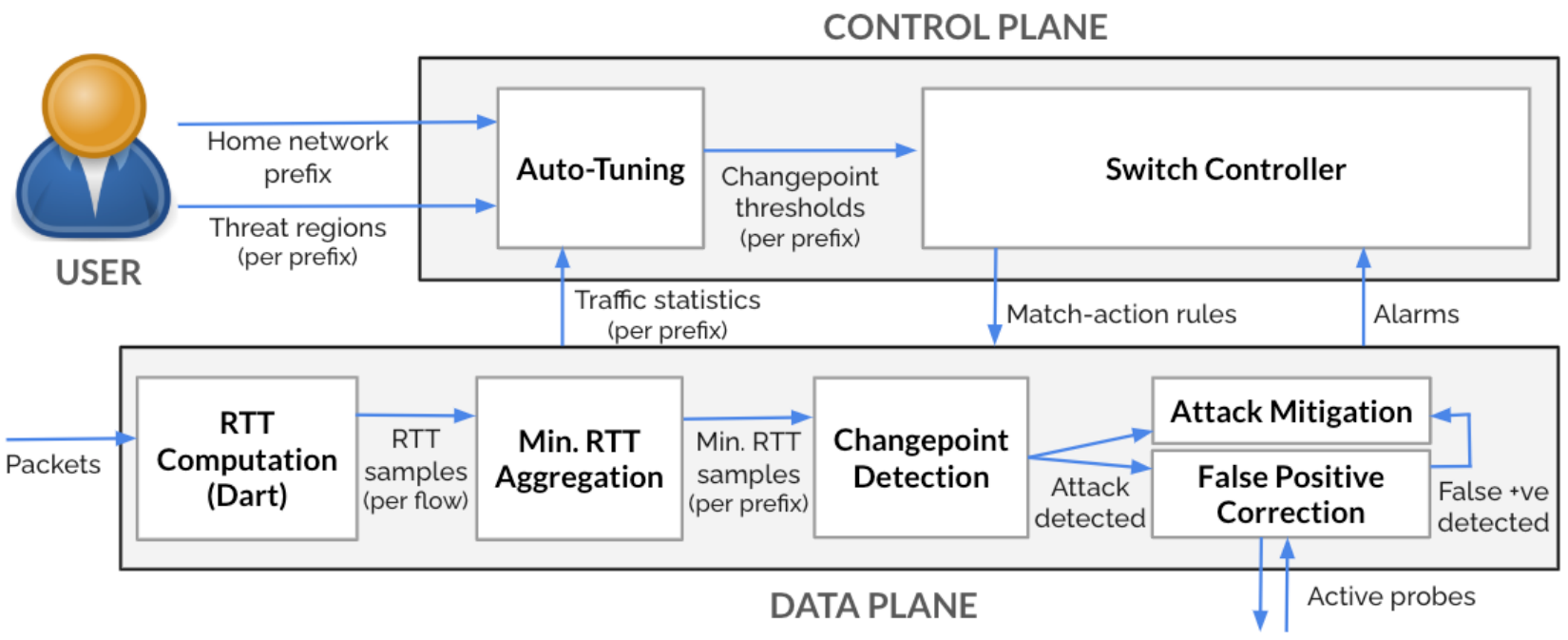}
\caption{\sys consists of a software control plane and a hardware data plane. The control plane auto-tunes per-prefix parameters for changepoint detection based on user inputs and traffic statistics from the data plane, and installs those parameters as match-action rules on the data plane. The data plane computes RTT samples, aggregates them by prefix, computes minRTT per window, and performs changepoint detection. Upon detecting an attack, the data plane blocks the corresponding prefix and triggers active probing to correct false positives.}
\label{fig:system-overview}
\end{figure*}

\noindent
Figure~\ref{fig:system-overview} presents an overview of \sys's end-to-end workflow. \sys comprises a control plane, implemented in software on a server, and a data plane, operating in high-speed hardware on a programmable switch. \sys is deployed at the edge of a production network, and can observe all or most of its traffic depending on the network topology (appendix~\ref{sec:deployment}).

\subsection{Control Plane}
\label{subsec:control-plane}

\boldpara{User Input.} The user configures the control plane by providing the network prefix of their home network (source prefix) and specifying the threat regions (optionally, per prefix).

\boldpara{Auto-Tuning.} The auto-tuning component translates the user inputs and destination prefixes read from the data plane into corresponding geographic coordinates and computes the $\tau_{mid}^*$.
It also retrieves traffic statistics (specifically, $min(RTT_{min})$ and $max(RTT_{min})$) from the data plane for each prefix. Combining this information, the component identifies which prefixes can be effectively protected and generates changepoint detection parameters for those prefixes, which it then sends to the switch controller. For prefixes that cannot be protected, it provides the user with a summary listing each prefix and their corresponding $RTT_{min}$ statistics.

\boldpara{Switch Controller.} The switch controller translates the received parameters into corresponding match-action rules and installs them on the data plane. It also receives attack alarms from the data plane and (optionally) notifies the user.

\subsection{Data Plane}
\label{subsec:data-plane}

\boldpara{RTT Computation.} We leverage Dart, an existing system, to generate accurate RTT measurements per flow from all the traffic observed by the switch. Dart achieves this at scale, handling a large number of flows without missing any RTT samples by efficiently managing switch resources~\cite{sengupta2022continuous}.

\boldpara{Min. RTT Aggregation.} The next component aggregates RTT samples per destination prefix, breaks them down into time windows,
and calculates the minimum RTT per window. Additionally, it computes traffic statistics like $min(RTT_{min}$ and $max(RTT_{min})$ per prefix to share with the control plane.

\boldpara{Changepoint.} The data plane performs changepoint detection using our two-window algorithm, minRTTs per window per prefix, and parameters installed by the control plane.

\boldpara{Attack Mitigation.} When an attack is detected, the data plane blocks the corresponding prefix and raises an alarm.

\boldpara{False Positive Correction.} The data plane then crafts and sends active probes periodically to determine whether the detection was a false positive. If so, it unblocks the prefix.

\subsection{Hardware Switch Prototype}
\label{sec:prototype}

\noindent
We implement our prototype in P4$_{16}$,
and deploy it on the Intel Tofino2 high-speed programmable switch, which supports up to 12.8 Tbps of traffic at line rate~\cite{budiu2017p416, agrawal2020intel}. Our prototype does not depend on any specific features available on the Tofino, and can be ported readily to other programmable packet-processing hardware including other switches (e.g., Juniper Trio) and SmartNICs (e.g., Nvidia BlueField3).

\boldpara{Switch control plane.}
The switch control plane installs per-prefix match-action rules specifying the window size ($W$), absolute threshold ($\tau_{mid}^*$), and surge threshold ($\lambda$). If enabled, it listens on the CPU port for packets from the data plane  containing information about either attack detections or non-coverage, and notifies the user. Additionally, it configures the Tofino's \emph{packet generator} to send active probes during the \emph{false positive correction} phase.

\boldpara{RTT computation.} We leverage Dart for continuous and accurate per-flow RTT computation~\cite{sengupta2022continuous}, and utilize \textit{packet mirroring}, a native feature that replicates packets, to enhance its functionality. First, the original packet is forwarded without added latency, with the mirrored copy used for RTT computation. Second, RTT samples generated by Dart are passed to \sys-specific data-plane components.

\boldpara{Per-prefix state.}
We maintain a \emph{prefix table} in register memory to support changepoint detection. The table uses a \emph{prefix signature}, derived by hashing the first 24 bits of the external IP, as the key. The stored values include the prefix's start timestamp, timestamp of most recent RTT, start timestamp of current window, number of RTT samples in current window, minRTTs for the current and previous windows, attack status,
$max(RTT_{min})$, and $min(RTT_{min})$. The table accommodates up to 65,536 active prefixes, significantly exceeding the peak observed in our 12h campus trace (approx. 5K assuming a 5-second timeout), minimizing hash collisions. For collisions, we use \emph{cuckoo hashing}~\cite{pagh2004cuckoo}: the new prefix replaces the old one, which is loaded into memory, checked for timeout, and recirculated to a new index using a different hash seed if still valid. Each insertion allows up to 3 recirculations.

\boldpara{Changepoint detection.}
When an RTT sample is generated for a prefix, \sys checks the status of the corresponding time window. If the window is not full, it updates the most recent timestamp, increments the RTT count, and replaces the current minimum RTT if the new sample is smaller. If the window is full, the current minimum RTT replaces the previous window's minimum, and $max(RTT_{min})$ and $min(RTT_{min})$ are updated. If the window is valid (i.e., it has enough samples), \sys evaluates the surge conditions and, if those are satisfied, starts the mitigation process by blocking all non-ICMP packets from/to the prefix by adding it to a \emph{block table}.

\boldpara{Active probing.} Simultaneously, we start crafting and sending ICMP echo packets to the latest IP seen from the prefix and listening to responses to monitor its RTT. If a false positive is detected, the prefix is removed from the block table.

\boldpara{Resource usage.} We analyze the resource usage of our prototype by function and find that its low resource consumption leaves ample resources for other concurrent switch functions (Table~\ref{tab:tofino-usage} in Appendix~\ref{sec:deployment}).
\section{Experimental Setup}
\label{sec:expt-setup}

\begin{figure}[t]
\centering
\includegraphics[width=.99\linewidth]{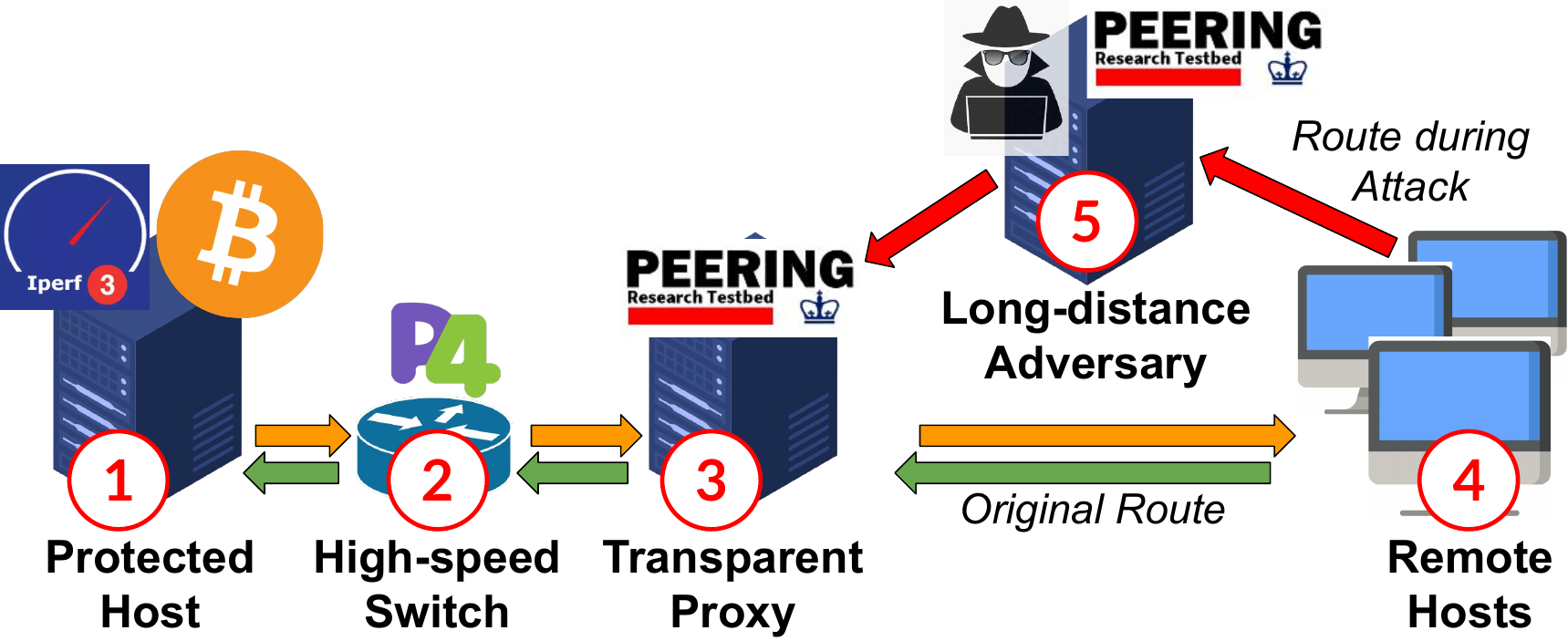}
\caption{Experimental setup for our live experiments. The orange and green arrows indicate the original ``protected host'' to ``remote hosts'' route and back, respectively. The return path (green) is intercepted by the \emph{long-distance adversary}---the diverted portion of the route is shown with red arrows.}
\label{fig:expt-system}
\end{figure}

\noindent
In this section, we outline our experimental setup: first, to demonstrate live detection of ethically launched interception attacks on controlled \emph{iperf} traffic; second, to showcase \sys's effectiveness \emph{in the wild} by detecting attacks on real \emph{Bitcoin} traffic; and third, to collect a 12-hour campus trace highlighting \sys's low false positive rate and minimal impact on regular operations.

\subsection{Passive Capture of Production Traffic}
\label{subsec:setup-passive}

\boldpara{Data collection.}
As outlined before, we captured 12h of production traffic---covering 1 pm to 1 am local time to include global working hours---at the edge of our US-based campus network using a TAP device near the gateway router. Packets---only TCP headers, anonymized at source in a prefix-preserved manner (Appendix~\ref{sec:ethics})---from selected subnets were mirrored and recorded on a collection server with \emph{tcpdump}.

\boldpara{Dataset overview.}
The dataset comprises $1.1$ TB of trace data representing $5.32$ TB of packet bytes, encompassing $19$ billion packets, $7.5$ million flows, and $238$ million RTT samples.
It includes $12$K unique internal IPs and $324$K unique external IPs, distributed across $183$K external prefixes, $23.2$K of which are based in the US.

\subsection{Live Experiments}
\label{subsec:setup-active}

\noindent
Figure~\ref{fig:expt-system} shows the experimental setup for our live experiments involving active traffic from the iperf3 and Bitcoin apps. Each component is marked with a number in red.

\boldpara{Deploying \sys to protect experimental traffic.} We set up our experiment using three key components: (1) a host on our campus running the application (\emph{iperf3} or \emph{Bitcoin}), (2) a high-speed programmable switch on campus where \sys{} is deployed to monitor traffic, and (3) a transparent TCP proxy on an Amazon AWS instance. The proxy, which doubles as a \emph{PEERING} node, advertises a /24 prefix allocated to our experiment. PEERING provides distributed ASes for controlled, real BGP announcements~\cite{schlinker19peering}. We use one IP address from the /24 pool, applying \emph{iptables} rules on components 1 and 3 to masquerade it as the application's IP. This setup enables \sys{} to monitor all experimental traffic while allowing external adversaries to ethically launch BGP interception attacks on the /24 prefix. The transparent proxy ensures the application on component 1 sees the remote host’s true IP, essential for applications like Bitcoin.

\boldpara{Setting up remote hosts.} For our experiments with \emph{iperf3}, we deploy AWS instances in geographically diverse locations, including the US east and west coasts, Europe, and Asia. The \emph{iperf3} server runs on the protected host across multiple ports, while clients on the distributed AWS instances connect to the PEERING IP of the transparent proxy, which forwards traffic to the server. For Bitcoin experiments, we run a node on the protected host, allowing random nodes worldwide to connect. Across different runs, we observe peers from the US, various European countries, and parts of Asia. We collectively refer to these remote hosts—used in both \emph{iperf3} and \emph{Bitcoin} experiments—as component 4.

\boldpara{Launching ethical routing attacks.} The final step in our setup is launching ethical BGP interception attacks on the PEERING IP. We designate the PEERING node in Amsterdam as the attacker (component 5) and implement a stealthy interception attack using the technique by Birge-Lee et al., which employs BGP communities to control the blast radius of the attack~\cite{birgelee2019sico}. The attacker advertises the same /24 prefix as the transparent proxy (an \emph{equally specific} attack), redirecting traffic from nearby nodes to Amsterdam. The attacker then forwards the intercepted traffic to the transparent proxy, leaving both sender and receiver unaware of the attack.
\section{Evaluation}
\label{sec:evaluation}

\noindent
In this section, we present our evaluation results. In the first part (Section~\ref{subsec:simulation-eval}), we run faithful simulations of \sys on our campus dataset and report coverage, false positive rate, and downtime due to false positives.
In the second part (Section~\ref{subsec:prototype-eval}), we present results from live experiments where the \sys{} prototype defends the protected host (in Figure~\ref{fig:expt-system}) when a subset of connections are impacted by an ethically conducted long-distance BGP interception attack.

\begin{figure*}
     \centering
     \begin{subfigure}[b]{0.23\textwidth}
         \centering
         \includegraphics[scale=0.2]{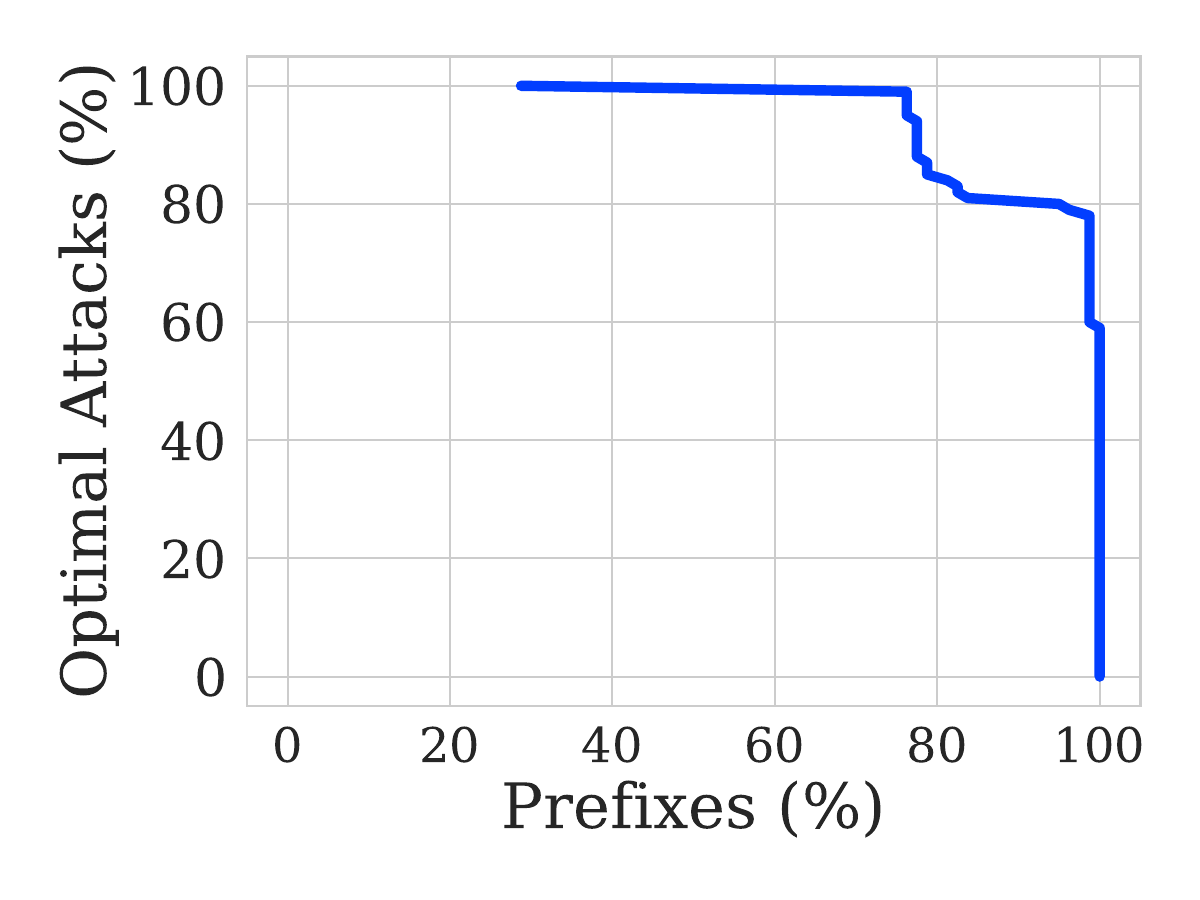}
         \caption{Coverage (lower bound)}
         \label{subfig:theory-coverage}
     \end{subfigure}
     \hfill
     \begin{subfigure}[b]{0.23\textwidth}
         \centering
         \includegraphics[scale=0.2]{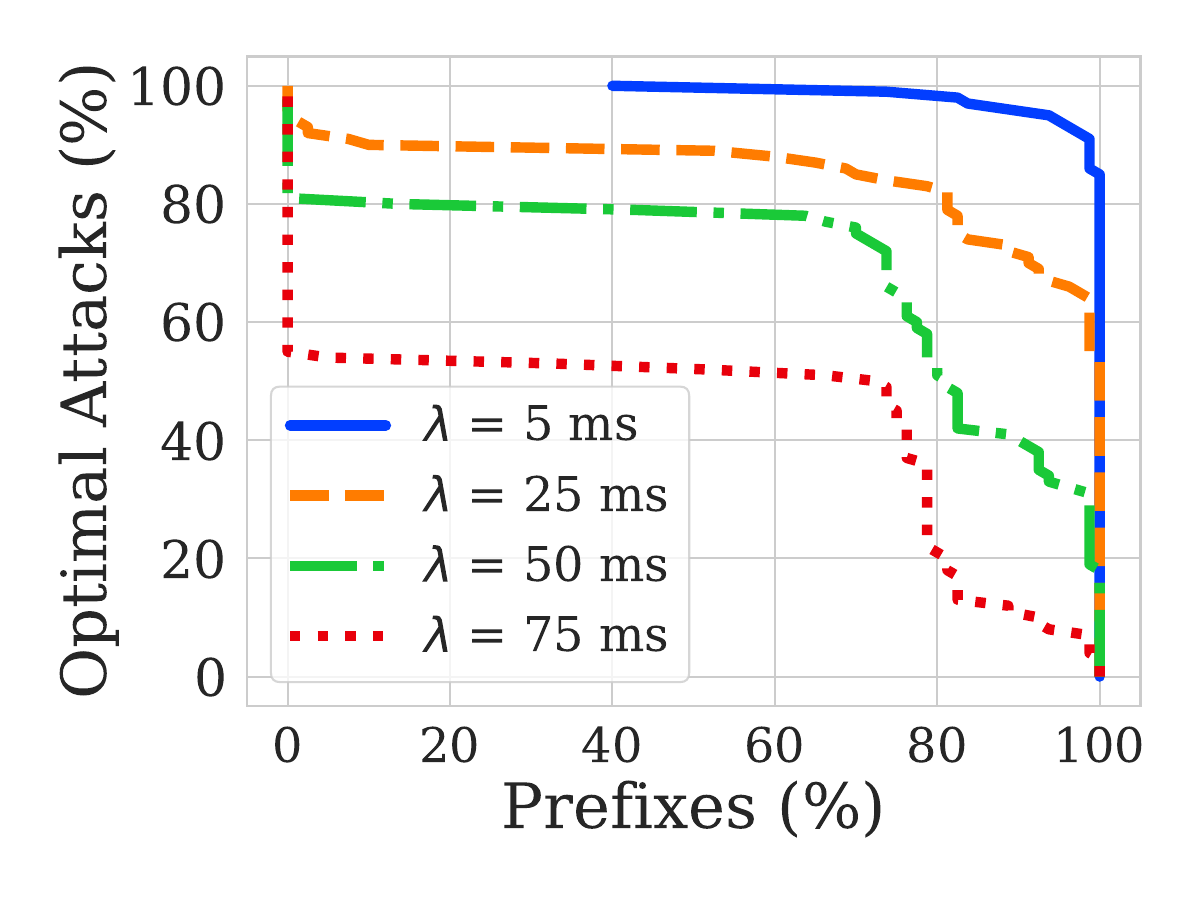}
         \caption{Coverage (defendability)}
         \label{subfig:defensability-coverage}
     \end{subfigure}
     \hfill
     \begin{subfigure}[b]{0.23\textwidth}
         \centering
         \includegraphics[scale=0.2]{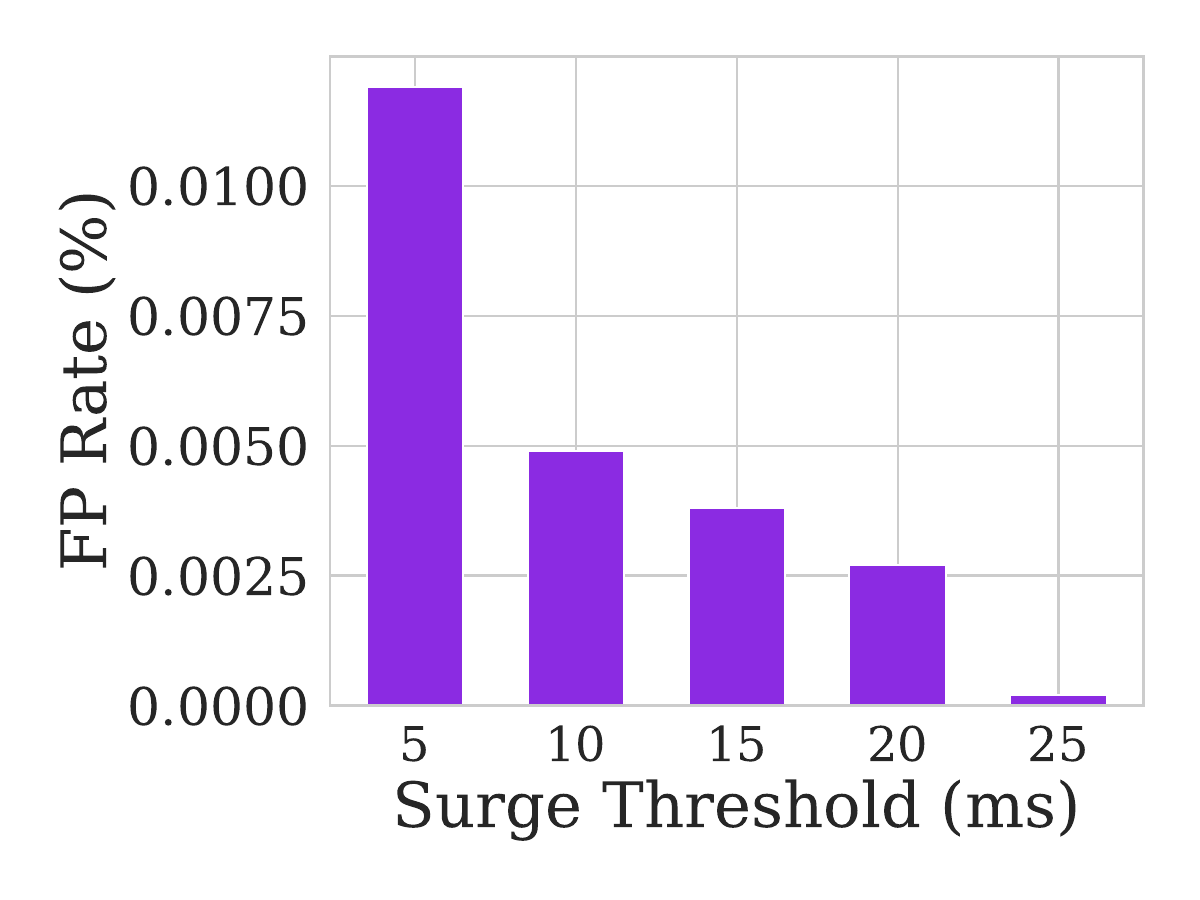}
         \caption{False positive rate}
         \label{subfig:false-positives}
     \end{subfigure}
     \hfill
     \begin{subfigure}[b]{0.23\textwidth}
         \centering
         \includegraphics[scale=0.2]{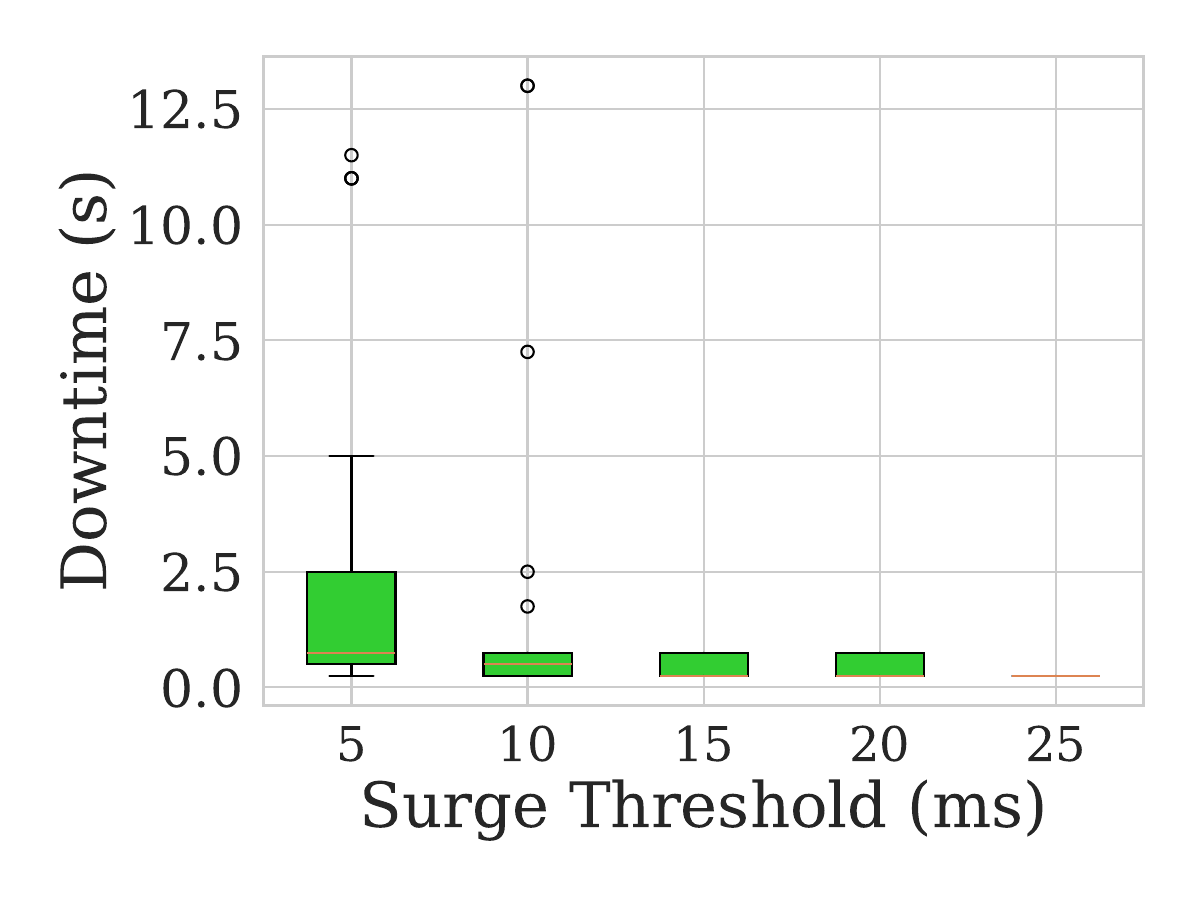}
         \caption{Downtime}
         \label{subfig:downtime}
     \end{subfigure}
        \caption{Faithful simulation on campus data illustrates that \sys can defend most prefixes from optimal attacks from most countries, incurs low false positives (<=0.012\%) and low downtime due to false positives (median<=0.75s).}
        \label{fig:simulation-results}
\end{figure*}

\subsection{Trace-based Evaluation}
\label{subsec:simulation-eval}

\noindent
We evaluate \sys using three metrics: (1) False positive rate or FPR (measures reliability/usability/practicality), (2) Coverage (measures the trade-off with low FNR and FPR), and (3) Downtime (measures impact of false positives on regular operation). We perform this evaluation using a faithful simulation of \sys written in Python on real latency data obtained from production traffic on our campus (Section~\ref{subsec:setup-passive}). We operate with the goal of protecting US-based prefixes from long-distance interception attacks from the mainlands of other countries in our dataset.
For each prefix, we divide into two equal parts the total time during which the prefix was active: the first half is used for \emph{profiling} while the second half is used for \emph{detection}.

\boldpara{Vulnerable prefixes.}
In accordance with \sys's coverage strategy, we only defend vulnerable prefixes, i.e., external prefixes associated with a server. We identify such prefixes using a TCP port number-based heuristic: external host's port < 1024 and internal host's port >= 1024. 16.8K US-based external prefixes match this condition.

\boldpara{Profiled prefixes.}
From external server prefixes, we further select those that were active for at least 10 minutes out of 12 hours (so we profile on at least 5 mins of data). We determine this by dividing the 12h period into buckets of 1 min, and checking which prefixes generated an RTT sample in at least 10 such buckets. 6K US-based external server prefixes are retained after this step.
In a real network, the operator could profile prefixes for as long as needed to ensure it covers typical variation of RTT during benign operation---without any excess overhead since the profiling happens directly in the switch. For the following analysis, we make the assumption that the campus data captured during the 12-hour period did not experience any long-distance interception attacks (i.e., no true positives were present), so if \sys detects a prefix it must be a false positive. We report our results based on optimal attacks from all 257 non-US threat countries.

\boldpara{Coverage impact of theoretical lower bound.}
Some US-based prefixes are not defendable against certain threat regions because during the profiling phase, they exhibit a $min(RTT_{min})$ larger than the corresponding $\tau_{mid}^*$ (i.e., measured delay without diversion is always higher than the lower bound with diversion). This could be because the threat region is geographically too close or because the network is always congested. Figure~\ref{subfig:theory-coverage} shows that, based on this condition, \sys can cover 99\% prefixes against 78\% attacks and 75\% prefixes against 99\% attacks. The covered prefix-threat country pairs are considered in the subsequent experiments.

\boldpara{Coverage impact of defensability analysis.}
For different values of the surge threshold ($\lambda$), Figure~\ref{subfig:defensability-coverage} shows \sys's coverage based on defendability---i.e., whether the mid-attack lower bound RTT clears the pre-attack $max(RTT_{min})$ by at least $\lambda$ ms. At $\lambda$ = 5 ms, 25 ms, 50 ms, and 75 ms respectively, \sys can cover 99\% prefixes against 91\%, 64\%, 31\%, and 7\% attacks, respectively. This illustrates the trade-off between surge threshold and coverage.

\boldpara{False positive rate.}
By focusing on defendable prefixes, we achieve a false positive rate of approximately $0.012\%$ at worst, as shown in Figure~\ref{subfig:false-positives}. For higher surge thresholds (i.e., 30 ms), the rate drops to zero.

\boldpara{Downtime due to false positives.} We estimate the likely downtime from a false positive by measuring how long (in multiples of time window size) it takes for the minRTT to return to normal for a falsely detected prefix. Since our active probing sends one probe per window, we expect similar results in reality. The median downtime is only 0.75 seconds.

\subsection{Live Interception Attack Detection}
\label{subsec:prototype-eval}

\begin{figure}
     \centering
     \begin{subfigure}[b]{0.47\textwidth}
         \centering
         \includegraphics[scale=0.22]{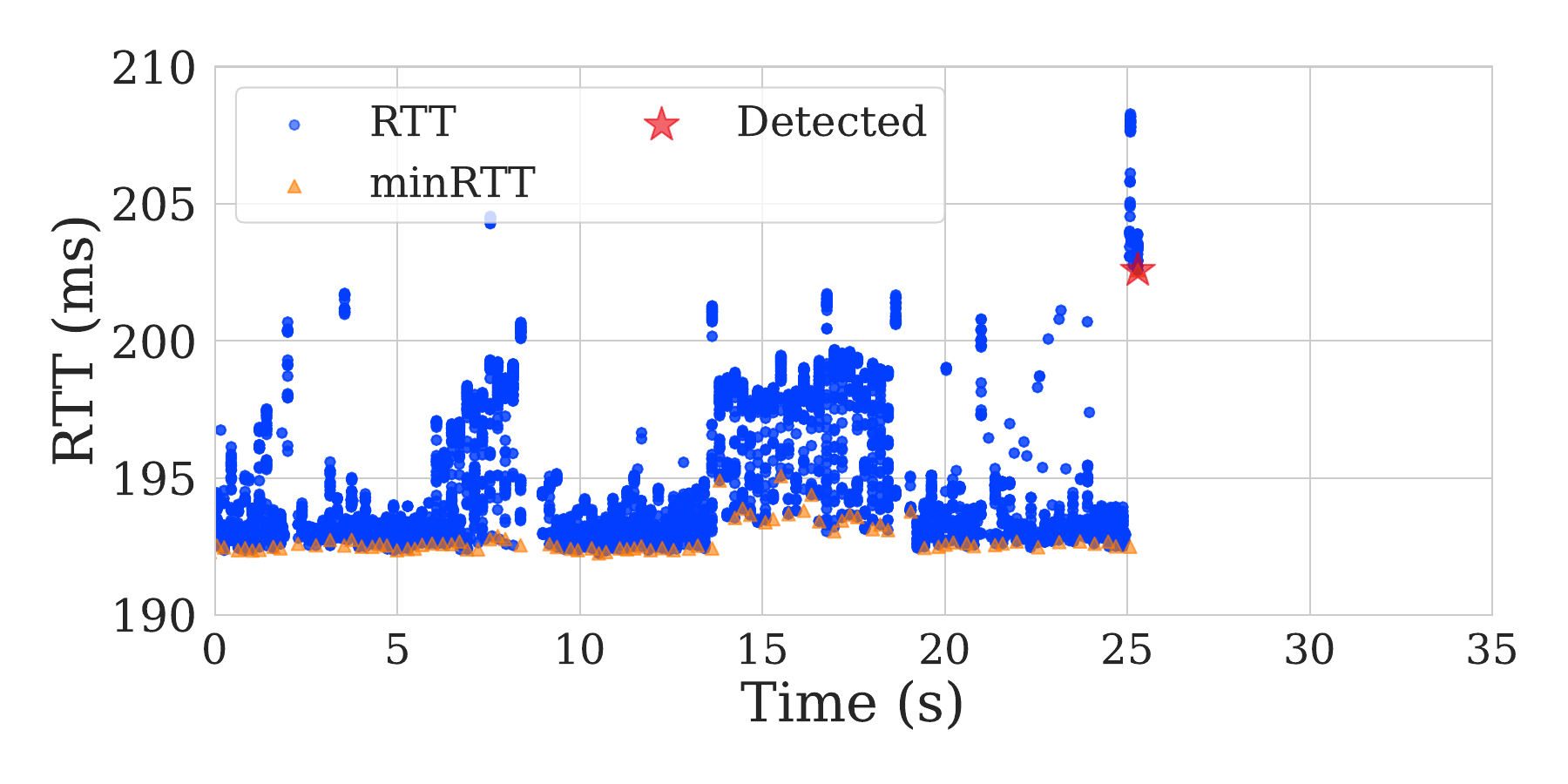}
         \caption{Detecting attack on iperf3 traffic.}
         \label{subfig:iperf-detection}
     \end{subfigure}
     \hfill
     \begin{subfigure}[b]{0.47\textwidth}
         \centering
         \includegraphics[scale=0.18]{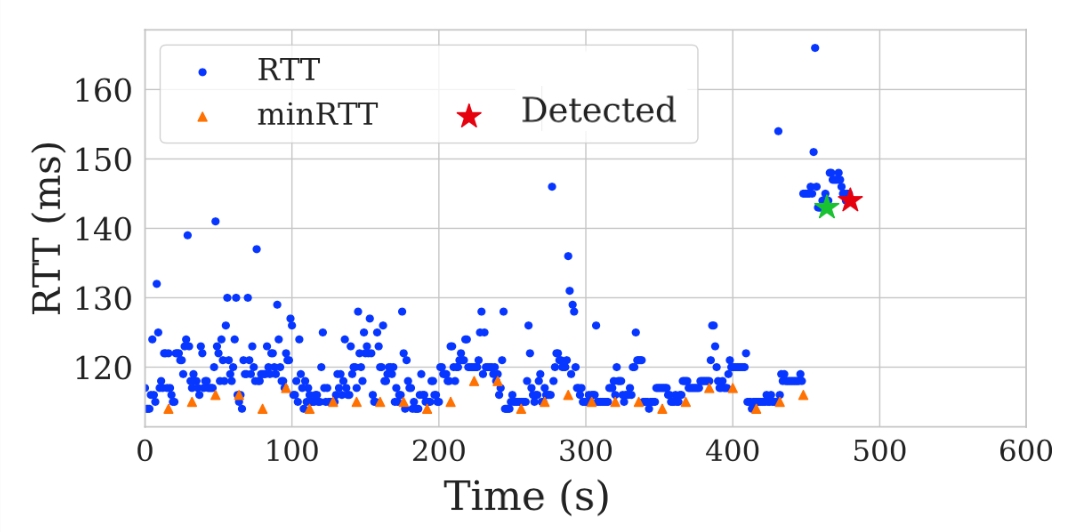}
         \caption{Detecting attack on Bitcoin traffic.}
         \label{subfig:bitcoin-detection}
     \end{subfigure}
    \caption{\sys (immediately) detects interception attacks ethically launched by us on iperf3 and Bitcoin traffic.}
    \label{fig:peering-detection}
\end{figure}

\boldpara{Mitigating attacks on \iperf traffic}:\\
\boldpara{Setup.} We run the iperf3 server on our campus and the transparent proxy in Ireland, who forwards all traffic it receives to our campus via our prototype. The iperf3 clients are in Virginia (2 flows from the same prefix), Ohio, and Mumbai. The prefix in Virginia is hijacked from Amsterdam, causing Ireland to send traffic to Amsterdam instead of Virginia. Amsterdam then forwards the traffic to Virginia. The traffic takes the following round-trip route before the attack: \emph{Virginia (via PEERING infra.) to Ireland to our campus to Ireland to Virginia (via PEERING infra.)} and the following one during the attack: Virginia (via PEERING infra.) to Ireland to our campus to Ireland to Amsterdam (via. PEERING infra.) to Virginia (via PEERING). Due to limitations of where we can deploy a Tofino switch on live traffic and a lack of diversity in the PEERING topology, we are restricted to this complex setup. The attack takes effect at $25$ seconds, as can be observed from the abrupt rise in RTT (blue dots) in Figure~\ref{subfig:iperf-detection}.

\boldpara{Interception detection.} Using multiple runs of traceroute, we estimate the lower bound of RTT as approx. 190.5 ms before attack and 199 ms during attack (absolute threshold). We set the window size to 0.25 sec and the surge threshold to 5 ms. Based on the minimum RTTs (orange triangles), we detect the attack almost immediately (red star).

\boldpara{Mitigating attacks on \emph{Bitcoin} traffic}:\\
Our setup is similar to the previous experiment with a Bitcoin application running on our campus being proxied to the host in Ireland. However, unlike the controlled iperf3 traffic in the previous case, real Bitcoin nodes from around the world---not controlled by us---connect to our application (ethical considerations are discussed in \S\ref{sec:ethics}).
We identify the nodes connecting from Europe using geolocation and install rules to monitor them. Then, we launch the interception attack ethically from Amsterdam. We set the absolute threshold to 135 ms based on our calculations and the surge threshold to 6 ms. Figure~\ref{subfig:bitcoin-detection} shows the effect of the attack on RTT samples followed by the detection and mitigation.
\section{Discussion}
\label{sec:discussion}

\boldpara{Faster data transmission.}
Certain networks, like free-space communication (e.g., microwave links) and satellite systems, can transmit data faster than speed of light in optical fiber. While not covered here, free-space networks are typically short-range and therefore irrelevant to long-distance rerouting, and satellite ASes can be easily identified by looking up their prefixes and excluding from our coverage.

\boldpara{Non-TCP traffic:} 
We do not need to explicitly monitor non-TCP flows (e.g., RTP or QUIC-over-UDP) to protect them from BGP interception attacks, provided there is at least one TCP flow within the same prefix generating RTT samples. These TCP flows, which share the same network path as the non-TCP flows, provide RTT measurements that can be used to detect potential attacks affecting the entire prefix.
\section{Related Work}
\label{sec:related-work}

\boldpara{Control-plane based detection:}
Control-plane approaches detect BGP hijacks by monitoring route advertisements~\cite{li2012buddyguard,sermpezis2018artemis,shi2012detecting}, but they are slow, with BGP convergence taking up to 30 seconds~\cite{holterbach2019blink,holterbach2017swift}. These methods can also be evaded by limiting route advertisements or targeting victims using BGP community manipulation~\cite{birgelee2019sico}.

\boldpara{Detection via active probing:}
Active probing methods use tools like ping, traceroute, and nmap to detect BGP hijacks. For example, iSPY detects hijacks in near real time by analyzing traceroutes from multiple vantage points~\cite{zhang2008ispy}. However, these approaches are vulnerable to surgical attacks~\cite{birgelee2019sico} and incur higher overhead compared to passive methods.

\boldpara{Detection via passive measurements:}
Passive measurements, such as RTT, have been used for BGP anomaly detection. Hiran et al. utilized crowd-sourced RTT data to detect attacks~\cite{hiran2015crowd}, though their method only addresses detection, not mitigation. Oscilloscope~\cite{buhler2023oscilloscope} offers advanced hijack detection but relies on emulated data, suffers from high false-positive and false-negative rates, and lacks data plane implementation, limiting its scalability.
\section{Conclusion}
\label{sec:conclusion}

\noindent
We present \sys, a system that detects and mitigates long-distance BGP interception attacks—where an adversary in another country hijacks traffic through its own infrastructure to eavesdrop before forwarding it to the victim. By leveraging propagation-delay measurements that attackers cannot conceal, \sys delivers high-accuracy defense at line rate on a programmable switch (Tbps). Our analysis of worst-case attacks across 258 countries confirms its effectiveness, and we validate \sys through simulations on anonymized campus traces and ethically conducted real-world hijacks, achieving robust mitigation with low false-positive rates.

\bibliographystyle{ACM-Reference-Format}
\bibliography{reference}

\appendix
\section{Ethics}
\label{sec:ethics}

This research study was reviewed and approved by our Institutional Review Board (IRB). All packet‐trace data come from our university network and were anonymized at the point of collection by network engineers who are expressly authorized to handle private data. Anonymization followed the exact procedures laid down by the IRB---anonymizing all IP and MAC addresses, and stripping all payloads. Researchers never had access to any raw or deanonymized data.

To validate \sys’s detection and mitigation capabilities in a live Internet environment, we performed two controlled BGP hijacks using prefixes assigned to us by the PEERING testbed \cite{peering}. We temporarily announced these prefixes from our own hosts under testbed guidelines, ensuring no impact on any external networks or clients. All BGP announcements and withdrawals adhered to PEERING’s guidelines, and only our own test prefixes were affected.

For our Bitcoin experiments, we operated a dedicated experimental node connected to public Bitcoin clients using Bitcoin's peer-to-peer protocol. We diverted only the traffic destined for our test node, maintaining the protocol’s standard multi-peer connectivity to ensure that no client experienced service interruption or security degradation. The increase in induced latency was temporary (under two minutes per peer) and well below the threshold required to compromise transaction privacy or network health.

\section{Open Science}
\label{sec:appendix:open-science}

We will release to the public---in a public GitHub repository---the source code of the \sys Tofino2 application written in P4, the complementary control plane written in Python, and the Jupyter notebooks used to conduct the analysis and evaluation presented in this paper. Additionally, we will also provide the exact steps to setup the \emph{Bitcoin} and \emph{iperf3} experiments used in this paper. For each component, we will also provide detailed documentation on how to install, deploy, and execute it. This work also utilizes \emph{anonymized} campus packet traces for part of the evaluation and analysis. These traces are protected by an Institutional Review Board (IRB) protocol and hence cannot be shared as-is. Instead, we will extract in a CSV file \emph{only the per-flow latency time-series} from these traces with the \emph{flow ID}, \emph{ACK timestamp}, and \emph{RTT} columns. The flow ID will be a unique identifier (e.g., f123) provided to each unique flow having no relation to the actual 5-tuple to preserve strict anonymity. We will release this CSV in the same repository to aid in the reproduction of the results and analysis presented in this paper.

We will publish all our code and materials in a public GitHub repository, including:
\begin{itemize}[leftmargin=*]
    \item The \sys Tofino2 P4 program;
    \item The complementary Python control‐plane code; and
    \item The Jupyter notebooks and C++ code used for our analysis and evaluation.
\end{itemize}
We will also include step-by-step instructions for setting up and running the Bitcoin and iperf3 experiments from the paper, along with full installation, deployment, and execution guides for every component.

Because our evaluation uses IRB‐protected, anonymized campus packet traces, we cannot share the raw data. Instead, we will provide a CSV containing only per‐flow RTT time series, with three columns: (1) Flow ID: a unique label (e.g., f123) unlinked to any actual 5-tuple, (2) ACK Timestamp, and (3) RTT in ms.
This CSV will appear alongside the code so that readers can reproduce our results without compromising privacy.

\section{Defendability based on speed-of-light RTTs}
\label{appendix:countries-cf}

\begin{figure}[t]
    \centering
    \includegraphics[scale=0.35]{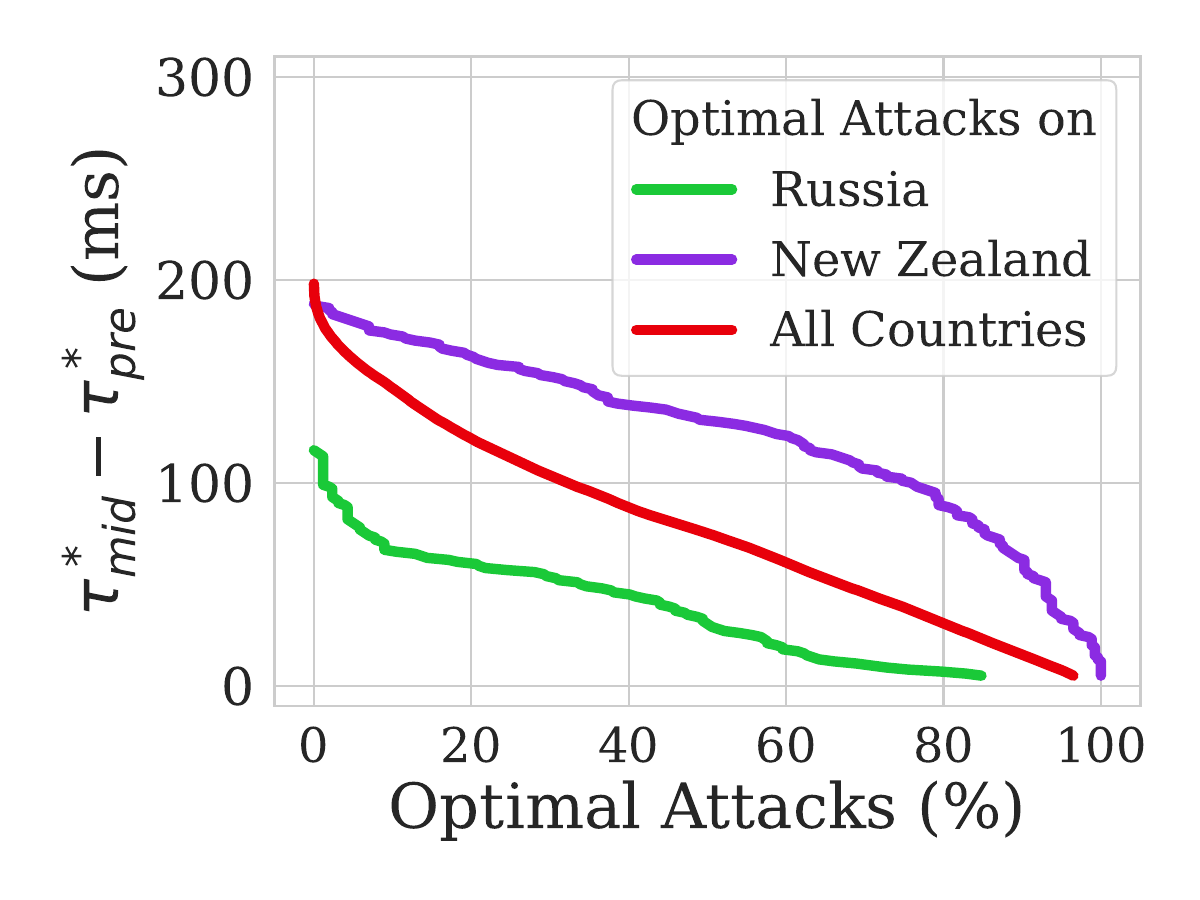}
    \caption{Attack coverage vs. minimum deviation (raw) for Russia, NZ, and all countries: With Russia and New Zealand (NZ) as example victim countries, the figure shows that the highest $\tau_{deviation}^*$ for Russia is 110 ms while it is 190 ms for NZ. For all countries combined, it is 200 ms.}
    \label{fig:mid_pre_diff}
\end{figure}

\noindent
Figure~\ref{fig:mid_pre_diff} shows the absolute difference between mid-attack and pre-attack RTTs vs. attack coverage in ideal conditions.

\section{Defendability based on measured RTTs}
\label{appendix:countries-measured}

\begin{figure}[t]
    \centering
    \includegraphics[scale=0.35]{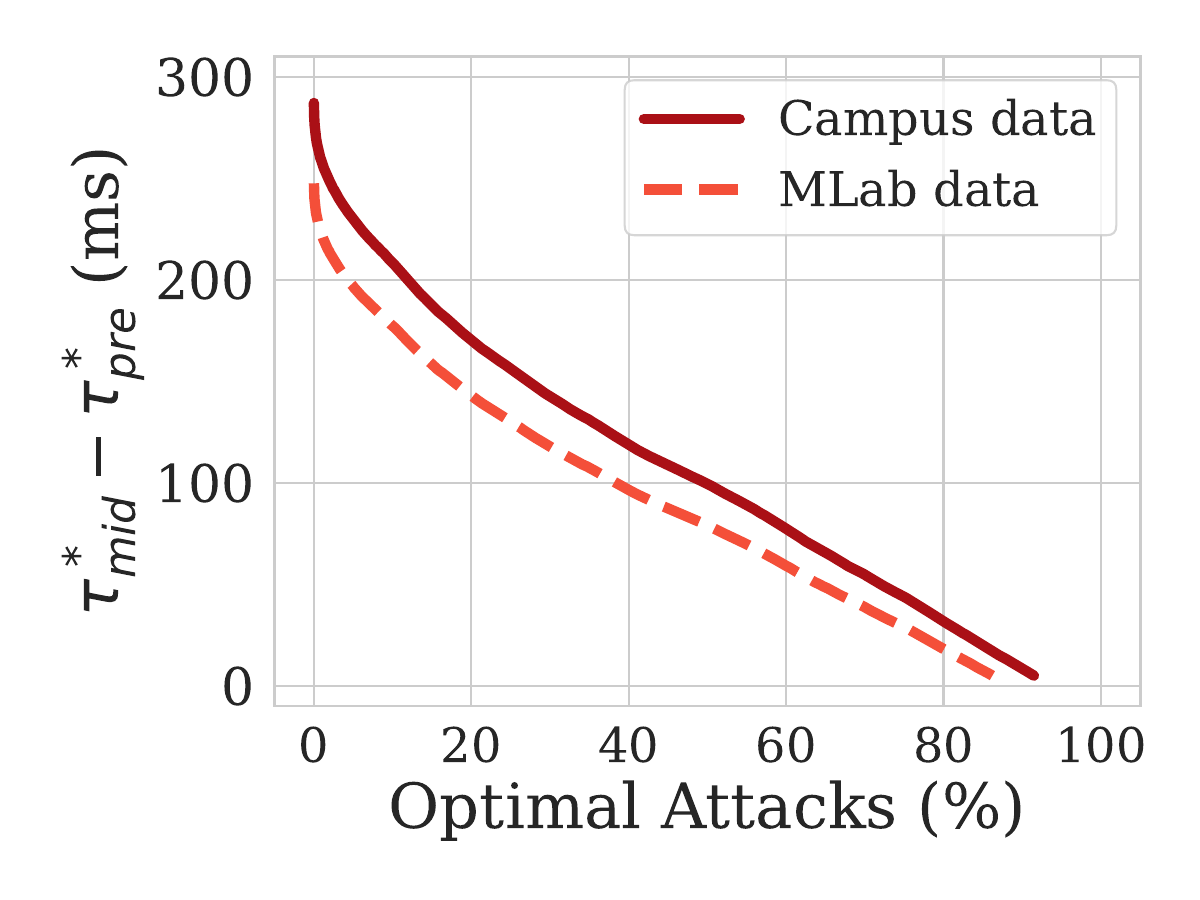}
    \caption{Attack coverage vs. minimum deviation (raw) for campus and MLab datasets.}
    \label{fig:cvg_diff_real}
\end{figure}

\begin{figure}[t]
    \centering
    \includegraphics[scale=0.35]{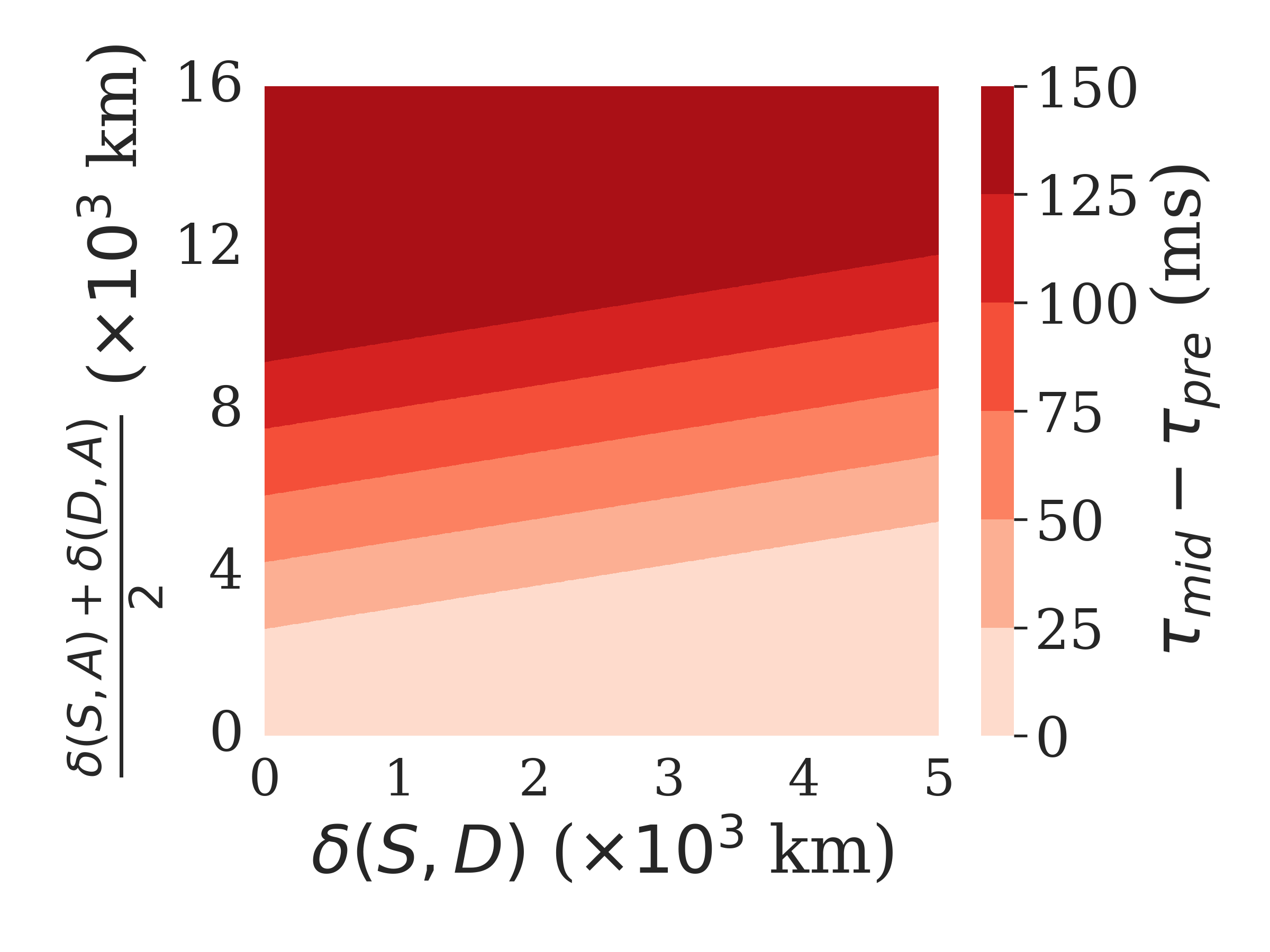}
    \caption{Campus dataset: Relationship between distances and minimum deviation.}
    \label{fig:campus_heatmap}
\end{figure}

\begin{figure}[t]
    \centering
    \includegraphics[scale=0.35]{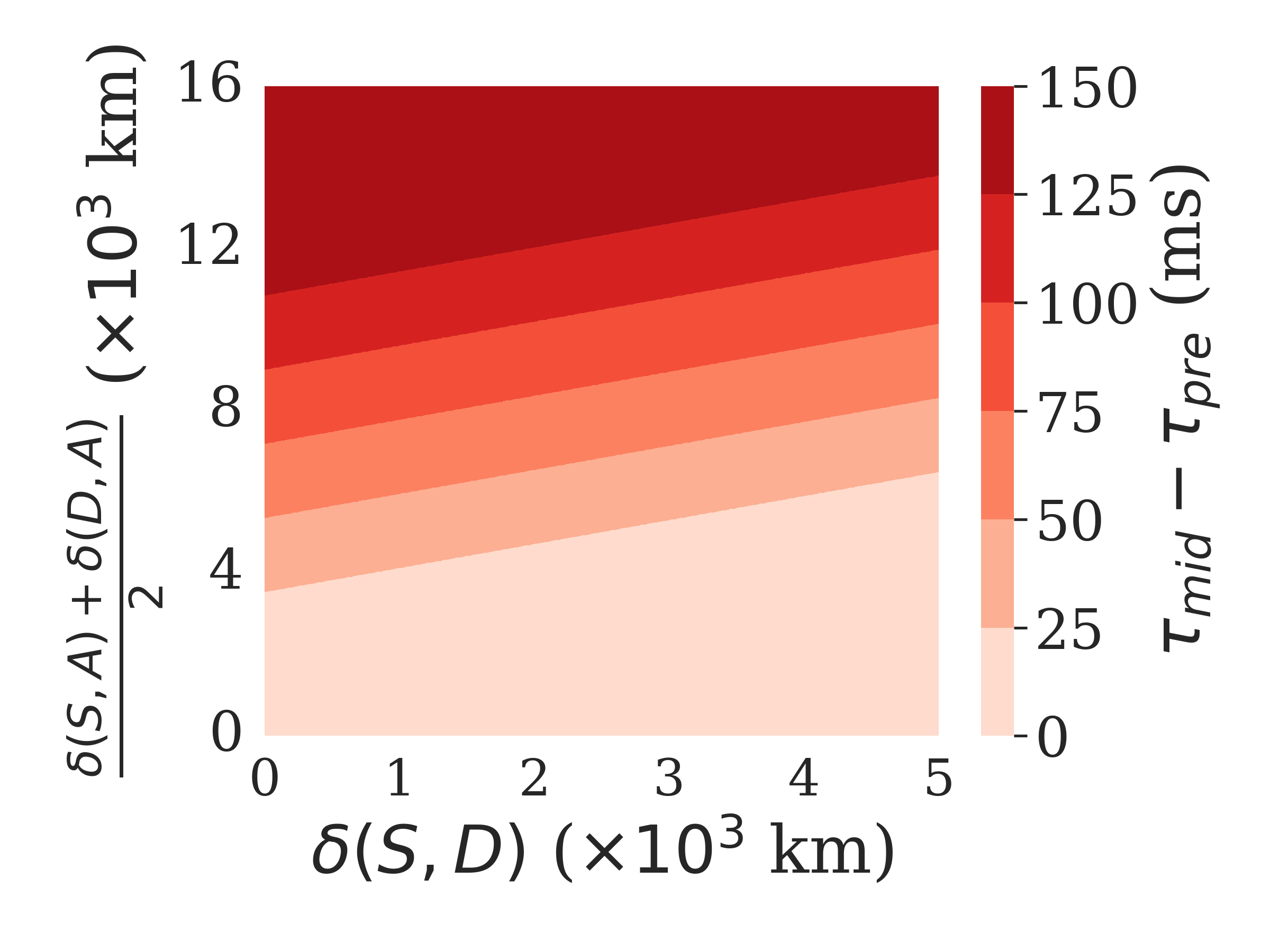}
    \caption{MLab dataset: Relationship between distances and minimum deviation.}
    \label{fig:mlab_heatmap}
\end{figure}

\noindent
Figure~\ref{fig:cvg_diff_real} shows the absolute difference between mid-attack and pre-attack RTTs vs. attack coverage in real-world conditions. Figures~\ref{fig:campus_heatmap} and~\ref{fig:mlab_heatmap} show the relationship between distance and minimum deviation under optimal attacks in the campus dataset and the MLab dataset, respectively.

\section{Deployment}
\label{sec:deployment}

\begin{figure}[t]
\centering
\includegraphics[width=.8\linewidth]{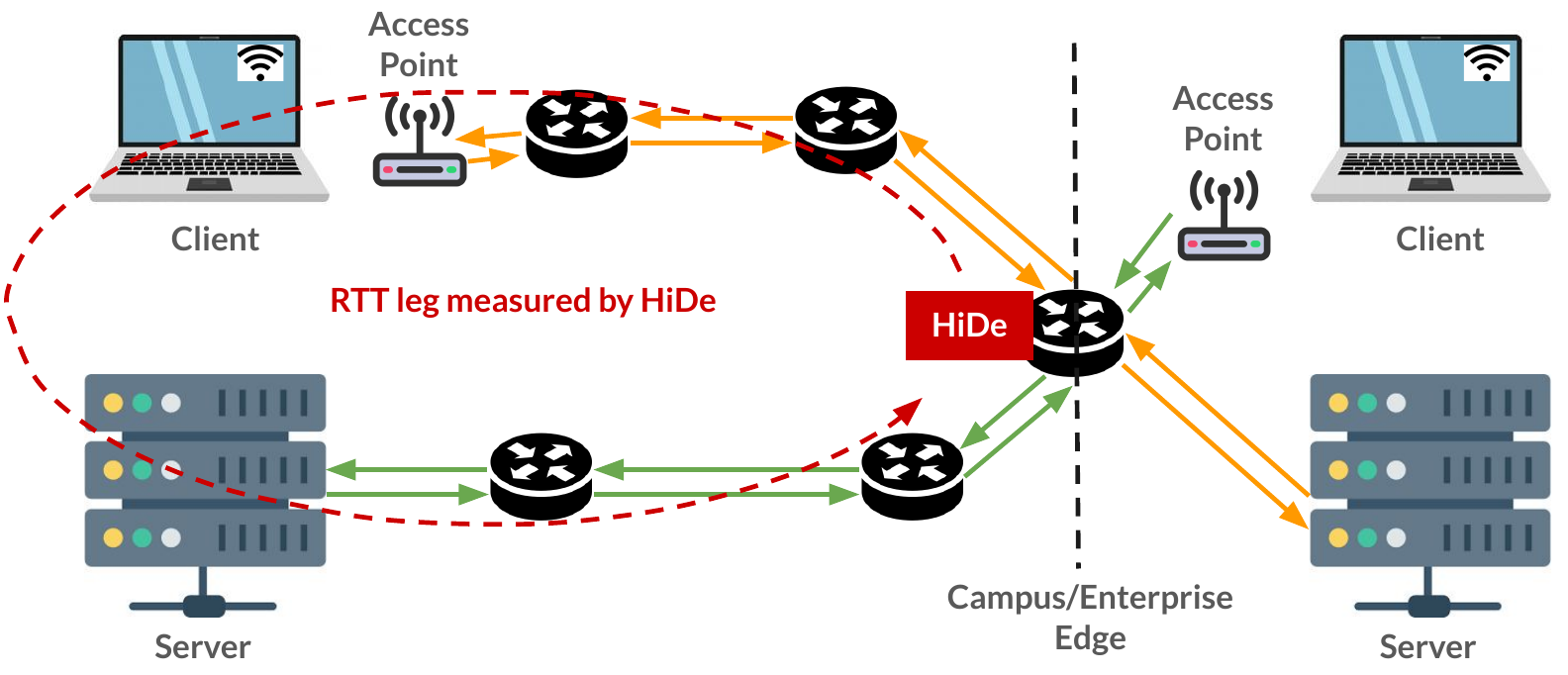}
\caption{\sys---deployed at the edge of a production network---defends servers and clients inside it by measuring the \emph{external leg} of RTT from itself to external hosts.}
\label{fig:deployment}
\end{figure}

\begin{table}[t]
    \small
    \centering
    \begin{tabular}{p{2cm} p{1.2cm} p{1.4cm} p{1.1cm} p{1.1cm}} 
        \textbf{Resource Type} & \textbf{Compute RTT~\cite{sengupta2022continuous}} & \textbf{Track Min. RTT} & \textbf{Detect Change} & \textbf{Mitigate Attack}\\ 
        \hline
        Stages & 7 & 2 & 4 & 3 \\
        TCAM & 2.9\% & 0.0\% & 1.1\% & 0.0\% \\
        SRAM & 4.5\% & 4.0\% & 2.4\% & 3.6\% \\
        Instructions & 3.6\% & 2.4\% & 1.0\% & 1.1\% \\
        Hash Units & 35.8\% & 12.5\% & 2.8\% & 5.6\% \\
        Input Crossbars & 10.1\% & 3.0\% & 1.6\% & 1.9\% \\
        \hline
    \end{tabular}
    \caption{Hardware resource usage of the Tofino2-based prototype, divided by functional component.}
    \label{tab:tofino-usage}
\end{table}

\noindent
\sys is deployed at the edge of a production network (Figure~\ref{fig:deployment}), protecting
clients within the network by monitoring the external leg of RTTs (\sys to external hosts) rather than the internal leg (\sys to internal hosts)~\cite{sengupta2022continuous}. We denote connections with clients inside the defended network as \emph{Client-In-Server-Out} (CISO) and connections with servers inside the network as \emph{Server-In-Client-Out} (SICO). We observe that the primary source of noise in RTTs is typically the access link near the client. For CISO connections, the access link is part of the internal leg and does not affect the monitored RTTs, resulting in less noise. In contrast, SICO connections experience higher noise levels, as the access link is external. In our campus data, we apply a TCP port number-based heuristic to distinguish CISO connections from SICO connections: if the port number used by the campus-internal host is $<1024$ and the one used by the external host is $>1024$, we consider it a SICO connection, and vice-versa.

\end{document}